\documentclass[12pt]{article}
\usepackage{amsfonts,amssymb,amsmath}
\usepackage{epic,eepic}
\usepackage{theorem,cite}
\newtheorem{definition}{Definition}[section]

\newtheorem{proposition}[definition]{Proposition}

\theorembodyfont{\rmfamily}

\numberwithin{equation}{section}

\def\cA{{\cal A}}          \def\cB{{\cal B}}          
          \def\cE{{\cal E}}          
\def\cG{{\cal G}}          \def\cH{{\cal H}}

\def\cP{{\cal P}}                    \def\cR{{\cal R}}
\def\cS{{\cal S}}          \def\cT{{\cal T}}          \def\cU{{\cal U}}
                    
\def\cY{{\cal Y}}          

\def\tK{{\widetilde {K}}}

\newcommand{\CC}{{\mathbb C}}
\newcommand{\II}{{\mathbb I}}

\newcommand{\ZZ}{{\mathbb Z}}
\newcommand{\eps}{{\varepsilon}}
\newcommand{\EE}{{\mathbb E}}

\newcommand{\atopn}[2]{\genfrac{}{}{0pt}{}{#1}{#2}}

\newcommand{\finproof}{{\hfill \rule{5pt}{5pt}}}
\newcommand{\cTbar}{\overline{\cT}}

\newcommand{\qmbox}[1]{{\qquad\mbox{#1}\quad}}
\def\qmbox#1{\qquad\mbox{#1}\quad}

\def\tr{\mathop{\rm Tr}\nolimits}
\def\str{\mathop{\rm Str}\nolimits}
\def\diag{\mathop{\rm diag}\nolimits}
\newcommand{\non}{\nonumber}
 \newcommand{\enne}{{\cal N}}
 \newcommand{\emme}{{\cal M}}
\newcommand{\eppe}{\widehat{{\cal P}}}

\begin{document}
\pagestyle{empty}

%%%%%%%%%%%%%%%%%%%%%%%%%%%%%%%%%
%%%%%  HEADINGS POUR DRAFT  %%%%%
%%%%%%%%%%%%%%%%%%%%%%%%%%%%%%%%%
% \markright{\today\dotfill DRAFT\dotfill }
% \pagestyle{myheadings}
\null
\vfill

\begin{center}

{\Large \textsf{General boundary conditions for the $sl(\enne)$
 and $sl({\emme}|\enne)$ \\[5mm]
  open spin chains}}

\vspace{10mm}

{\large D. Arnaudon$^a$\footnote{arnaudon@lapp.in2p3.fr,
  avan@ptm.u-cergy.fr, crampe@lapp.in2p3.fr,\\ \null\qquad
  doikou@lapp.in2p3.fr,
  frappat@lapp.in2p3.fr, ragoucy@lapp.in2p3.fr\label{foot:1}},
  J. Avan$^{b\ref{foot:1}}$,  
  N.~Cramp{\'e}$^{a\ref{foot:1}}$,
  A.~Doikou$^{a\ref{foot:1}}$, L. Frappat$^{ac\ref{foot:1}}$,
  {E}. Ragoucy$^{a\ref{foot:1}}$} 

\vspace{10mm}

\emph{$^a$ Laboratoire d'Annecy-le-Vieux de Physique Th{\'e}orique}

\emph{LAPTH, CNRS, UMR 5108, Universit{\'e} de Savoie}

\emph{B.P. 110, F-74941 Annecy-le-Vieux Cedex, France}

\vspace{7mm}

\emph{$^b$ Laboratoire de Physique Th{\'e}orique et Mod{\'e}lisation}

\emph{Universit{\'e} de Cergy, 5 mail Gay-Lussac, Neuville-sur-Oise}

\emph{F-95031 Cergy-Pontoise Cedex}

\vspace{7mm}

\emph{$^c$ Member of Institut Universitaire de France}

\end{center}

\vfill

\begin{abstract}
Two types of boundary conditions (`soliton preserving' and
`soliton non-preserving') are investigated for the $sl(\enne)$ and 
$sl({\emme}|{\enne})$ open spin chains. 
The appropriate reflection equations are
formulated and the corresponding solutions are classified. The symmetry and
the Bethe Ansatz equations are derived for each case. 

The general treatment for non-diagonal reflection matrices associated 
to `soliton preserving' case is worked out.
The connection
between the `soliton 
non-preserving' boundary conditions and the twisted (super) Yangians is
also discussed.
\end{abstract}

\vfill
\vfill
\vfill

\rightline{math-ph/0406021}
\rightline{LAPTH-1050/04}
\rightline{June 2004}

\baselineskip=16pt

%%%%%%%%%%%%%%%%%%%%%%%%%%%%%%%%%%%%%%%%%%%%%%%%%%%%%%%%%%%%%%%%%%%%%%%%%%%%%%%
\newpage

\pagestyle{plain}
\setcounter{page}{1}
\section*{Introduction}
The possibility of constructing and (at least partially) solving by
algebraic and/or analytical methods, one-dimensional interacting quantum
spin chains, is one of the major achievements in the domain of quantum
integrable systems. Its main tool is the quantum $R$-matrix, obeying a
cubic Yang-Baxter equation, the ``coproduct'' properties of which allow the
building of an $L$-site transfer matrix with identical exchange
relations and the subsequent derivation of
quantum commuting Hamiltonians \cite{baxter}. The same structure is
instrumental in formulating the quantum inverse scattering procedure,
initiated by the Leningrad school \cite{FRT}.

A subsequent development was the definition of exactly solvable open spin
chains with non-trivial boundary conditions. These are characterised by a
second object: the reflection matrix $K$, obeying a quadratic consistency
equation with the $R$ matrix, with the generic abstract form $RKRK = KRKR$
\cite{cherednik,sklyanin,KuSk,KuSa,DoKuMu}. Using again ``coproduct-like''
properties of this structure one 
constructs suitable transfer matrices yielding (local) commuting spin chain
Hamiltonians by combining $K$ and semi-tensor products of $R$
\cite{sklyanin}. 

Many efforts have been devoted to this issue
\cite{dvgr,GZ,abad,done,boco,gand,doikou1,deve,doikou2,dewo,saweka,%
yabon,banania,GR,Essler,referee}, 
based on the pioneering approach of Sklyanin 
and we here aim at treating a particular,
but very significant, class of examples for this problem.

To better characterise the type of spin chain which we will be considering
here it is important to recall that both $R$ and $K$ matrices have an
interpretation in terms of diffusion theory for particle-like objects
identified in several explicit cases with exact eigenstates of some quantum
integrable field theories such as sine--Gordon\cite{zamo},  non-linear
Schr\"odinger equation \cite{sklyanin,GLM} or 
principal chiral model \cite{McS}. $R$ describes the basic $2$-body
scattering amplitudes and $K$ describes a $1$-body scattering or reflection
on a boundary. The Yang-Baxter equation (YBE) and reflection equation (RE)
then characterise consistent factorisability of any $L$-body amplitude in
terms of $1$- and $2$-body scattering amplitudes, regardless of the order of
occurrence of the $1$- and $2$-body events in the diffusion process.

As a consequence, when one describes the scattering theory of a model with
more than one type of particle involved, one is led to introduce several
operators of $R$ and $K$ type. The case which we examine here corresponds to
a situation where the states involved can be split into resp. particles and
anti-particles with a suitable representation of $CP$ transformation acting
on the states. Within the context of integrable field theories it is
justified to denote them respectively ``solitons'' $S$ and ``antisolitons''
$A$. Assuming that the $2$-body diffusion conserves the soliton or
antisoliton nature of the particles, but that the reflection may change it,
one should therefore consider four types of $R$ matrices (resp.
$R^{SS}_{SS}$, $R^{SA}_{AS}$, $R^{AS}_{SA}$, $R^{AA}_{AA}$) connected by
$CP$ operations; and four reflection matrices (resp. $K_{S}^{S}$,
$K_{A}^{S}$, $K_{S}^{A}$, $K_{A}^{A}$).
It then becomes possible to define several non-equivalent constructions of
commuting transfer matrices. As a consequence, one sees that a
variety of spin chain models can be built using a Sklyanin-like procedure,
depending on which transfer matrix is being constructed and which
reflection matrices are used to build it. Locality arguments also come into
play, leading to more complicated combinations of transfer matrices as we
shall presently see.

We shall here describe the construction, and present the resolution by
analytical Bethe Ansatz methods \cite{mnanal,doikou2}, of open spin 
chains based on
the simplest rational $R$ matrix solutions of the Yang-Baxter equations for
underlying $sl(\enne)$ and $sl(\emme|\enne)$ Lie (super) algebras. These
solutions, with a rational dependence on the spectral parameter, are
instrumental in defining the Yangian \cite{Drinfeld}. We shall consider
two types of 
associated reflection matrices $K$ to build two distinct types of
integrable spin chains: one which entails purely soliton- and
antisoliton-preserving reflection amplitude by two matrices $K_{S}^{S}$ and
$K_{A}^{A}$ (hereafter denoted ``soliton-preserving case'' or SP); the
other which entails the two soliton-non-preserving reflection amplitudes
$K_{S}^{A}$ and $K_{A}^{S}$ (hereafter denoted ``soliton-non-preserving
case'' or SNP).
Closed spin chains based on $sl(\emme|\enne)$ superalgebras were
studied in e.g. \cite{saleur1} and, in the case of alternating
fundamental-conjugate  
representations of $sl(\emme|\enne)$ in \cite{mmartins}.
Open spin chains based on $sl(1|2)$ have been studied in details in
e.g. \cite{GR,Essler}.

The plan of our presentation is as follows:

We first define the relevant algebraic objects, $R$ matrix and $K$ matrix,
together with their compatibility (Yang--Baxter and reflection) equations
and their relevant properties. In particular we introduce the various
reflection equations which arise in the SP and SNP cases. Let us emphasise
that all notions introduced in the $sl(\enne)$ case will be straightforwardly
generalised to the $sl(\emme|\enne)$ case, albeit with a graded tensor
product.

In a second part we define
the commuting transfer matrices which can be built in both cases, and the
local Hamiltonians which can be built from them. Locality requirement leads
to considering a product of two transfer matrices, resp.
soliton-antisoliton and antisoliton-soliton, in the SNP case.
Once again this construction will be valid, with suitable
modifications, for the $sl(\emme|\enne)$ case.

In a third part we discuss the symmetries of these transfer matrices
induced by their respective YBE and RE structures, in particular focusing
on the connection between the SNP case and twisted Yangians.

We then start the discussion of the analytical Bethe Ansatz formulation for
the $sl(\enne)$ spin chains in the SNP case. The derivation of suitable new
fusion formulae explicited in Appendix A and B makes it possible to get a
set of Bethe equations.

In Section 5 we consider the case of $sl(\emme|\enne)$ super algebra as
underlying algebra. Contrary to the previous case it is first needed to
establish a classification for the reflection matrices based on the
rational (super Yangian) quantum $R$-matrix solution, both for SP and SNP
conditions. We then establish the Bethe equations for both SP and SNP
cases. In the SP case in addition we consider spin chains built from
general $K$ matrix solutions, in the SNP case we restrict ourselves to
diagonal $K$ matrices.

\section{Yang--Baxter and reflection equations}

The $R$ and $K$ matrices obey sets of coupled consistency
equations together with characteristic properties which we
now describe. 

\subsection{The $R$ matrix\label{sect:Rmatrix}}

We will consider in a first stage the $sl(\enne)$ invariant $R$ matrices
\begin{eqnarray}
  R_{12}(\lambda) = \lambda \II+i\cP_{12}
\label{r}
\end{eqnarray}
where $\cP$ is the permutation operator
\begin{equation}
  \label{eq:P12}
  \cP_{12} = \sum_{i,j=1}^\enne E_{ij} \otimes E_{ji} \;.
\end{equation}
$E_{ij}$ are the elementary matrices with 1 in position $(i,j)$ and 0
elsewhere. 
\\
We define a transposition $^t$ which is related to the usual
transposition $^T$ by 
($A$ is any matrix):
\begin{equation}
A^t=V^{-1}\,A^T\,V \qmbox{where}\left\{
\begin{array}{ll}
  V = \mbox{antidiag}(1,1,\ldots,1)\,, &\ \mbox{ for which }\
  V^2=\theta_0=1\\
  \mbox{or}&\\
  V=\mbox{antidiag}\Big(\,
  \underbrace{1,\ldots,1}_{\enne/2}\,,\,\underbrace{-1,\ldots,-1}_{\enne/2}\,
    \Big)\,,
    &\ \mbox{ for which }\ 
  V^2=\theta_0=-1\,.
\end{array}\right.
\label{eq:V}
\end{equation}
The second case is forbidden for $\enne$ odd.\\
This $R$ matrix satisfies the
following properties: \\[2mm] 
\textit{(i) Yang--Baxter equation} \cite{mac,yang,baxter,korepin}
\begin{eqnarray}
  R_{12}(\lambda_{1}-\lambda_{2})\ R_{13}(\lambda_{1})\ R_{23}(\lambda_{2})
  =R_{23}(\lambda_{2})\ R_{13}(\lambda_{1})\ R_{12}(\lambda_{1}-\lambda_{2})
  \label{YBE}
\end{eqnarray}
\textit{(ii) Unitarity}
\begin{eqnarray}
  R_{12}(\lambda)\ R_{21}(-\lambda) = \zeta(\lambda) \label{uni1}
\end{eqnarray}
where $R_{21}(\lambda) =\cP_{12} R_{12}(\lambda) \cP_{12} =
R_{12}^{t_{1}t_{2}}(\lambda) = R_{12}(\lambda)$.
\\[2mm]
\textit{(iii) Crossing-unitarity}
\begin{eqnarray}
  R_{12}^{t_{1}}(\lambda)\ R_{12}^{t_{2}}(-\lambda-2i\rho)\
  = \bar\zeta(\lambda + i\rho)
  \label{croun}
\end{eqnarray}
where $\rho = {\enne\over 2}$ and
\begin{eqnarray}
  \zeta(\lambda) = (\lambda+i)(-\lambda +i), 
  \qquad
  \bar\zeta(\lambda) =
  (\lambda+i\rho)(-\lambda +i\rho).
\end{eqnarray}
It obeys 
\begin{equation}
{[A_{1} A_{2},\ R_{12}(\lambda)]} = 0 \qmbox{for any matrix $A$.} 
\label{RAA}
\end{equation}

The $R$ matrix can be interpreted physically as a scattering matrix
\cite{zamo, korepin, faddeev} describing the interaction between two
solitons that carry the fundamental representation of
$sl(\enne)$.

To take into account the existence, in the general case, of anti-solitons
carrying the conjugate representation of $sl(\enne)$, we shall introduce
another scattering matrix, which describes the interaction between a soliton
and an anti-soliton. This matrix is derived as follows
\begin{eqnarray}
  R_{\bar 12}(\lambda) =
  \bar R_{12}(\lambda) &:=& 
  R_{12}^{t_{1}}(-\lambda -i\rho)
  \label{cross}
  \\
  &=&
  R_{12}^{t_{2}}(-\lambda -i\rho)  
  \;=:\; 
  R_{1 \bar 2}(\lambda) 
  = \bar R_{21} (\lambda)
  \label{eq:br}
\end{eqnarray}
In the case $\enne=2$ and for $\theta_0=-1$ ($sp(2)$ case), $\bar R$
is proportional to 
$R$, so that there is no genuine notion of anti-soliton. This reflects the
fact that the fundamental representation of $sp(2)=sl(2)$ is self-conjugate.
This does not contradict the fact that for $\enne=2$ and for
$\theta_0=+1$ ($so(2)$ case), there exists a notion of soliton and
anti-soliton. 

The equality between $R_{\bar 12}(\lambda)$
and $R_{1 \bar 2}(\lambda)$  in (\ref{cross}) reflects the CP invariance of
$R$, from which one also has $R_{\bar 1 \bar 2} = R_{12}$, i.e. the
scattering matrix of anti-solitons is equal to the scattering matrix of
solitons. In (\ref{cross}), $\bar R_{12}(\lambda) = (-\lambda
-i\rho)\II+iQ_{12},$ in which $Q_{12}$ is proportional to a projector onto
a one-dimensional space. It satisfies
\begin{eqnarray}
  Q^{2} = 2 \rho\; Q\qquad \mbox{and} 
  \qquad 
  \cP\ Q =Q\ \cP = \theta_0 Q \;.
\label{proj}
\end{eqnarray}

The $\bar R$ matrix (\ref{eq:br}) also obeys
\\[2mm]
\textit{(i) A Yang--Baxter equation}
\begin{eqnarray}
  \bar R_{12}(\lambda_{1}-\lambda_{2})\ \bar R_{13}(\lambda_{1})\
  R_{23}(\lambda_{2}) =R_{23}(\lambda_{2})\ \bar R_{13}(\lambda_{1})\ \bar
  R_{12}(\lambda_{1}-\lambda_{2})
\label{YBE2}
\end{eqnarray}
\textit{(ii) Unitarity}
\begin{eqnarray}
  \bar R_{12}(\lambda)\ \bar R_{2 1}(-\lambda) = \bar\zeta(\lambda) 
\label{uni2}
\end{eqnarray}
\textit{(iii) Crossing-unitarity}
\begin{eqnarray}
  \bar R_{12}^{t_{1}}(\lambda)\ \bar
  R_{12}^{t_{2}}(-\lambda-2i\rho)\ =\zeta(\lambda+i\rho) \;.
\label{croun2}
\end{eqnarray}

\paragraph{Remark:}
The crossing-unitarity relation written in the literature usually involves a
matrix $M=V^T V$. In our case, $M$ turns out to be 1 for two reasons: (i) the
factors $q^k$ of the quantum (trigonometric) case degenerate to 1 in the
Yangian (rational) case; (ii) the signs usually involved in the super case
are in this paper (section \ref{sect:slmn}) taken into account in the
definition of the super-transposition (\ref{eq:st}).

\subsection{The $K$ matrix}
\label{sect:Kmatrix}

The second basic ingredient to construct the open spin chain is the $K$
matrix. We shall describe in what follows two different types of boundary
conditions, called \textit{soliton preserving} (SP)
\cite{dvgr,GZ,abad,done} and \textit{soliton non--preserving} (SNP)
\cite{boco,gand,doikou1}. \\

\subsubsection{Soliton preserving reflection matrices}

In the case of soliton preserving boundary conditions, the matrix
$K$ is a numerical solution of the reflection (boundary Yang--Baxter) 
equation
\cite{cherednik}
\begin{equation}
  R_{ab}(\lambda_{a}-\lambda_{b})\ K_{a}(\lambda_{a})\
  R_{ba}(\lambda_{a}+\lambda_{b})\ K_{b}(\lambda_{b})=
  K_{b}(\lambda_{b})\
  R_{ab}(\lambda_{a}+\lambda_{b})\ K_{a}(\lambda_{a})\
  R_{ba}(\lambda_{a}-\lambda_{b}) \;,
  \label{re}
\end{equation}
and it describes the reflection of a soliton on the boundary, coming
back as a soliton.

Another reflection equation is required for what follows, in particular
for the `fusion' procedure described in the appendices
\begin{equation}
  \bar R_{ab}(\lambda_{a}-\lambda_{b})\ K_{\bar a}(\lambda_{a})\ \bar
  R_{ba}(\lambda_{a}+\lambda_{b})\ K_{b}(\lambda_{b})= K_{b}(\lambda_{b})\
  \bar R_{ab}(\lambda_{a}+\lambda_{b})\ K_{\bar a}(\lambda_{a})\ \bar
  R_{ba}(\lambda_{a}-\lambda_{b}) \;.
  \label{re3}
\end{equation}
$K_{\bar a}$ is a solution of the anti-soliton reflection equation obtained
from (\ref{re}) by CP conjugation and actually identical to (\ref{re}) due
to the CP invariance of the $R$-matrix. It describes the reflection of an
anti-soliton on the boundary, coming back as an anti-soliton.

Equation (\ref{re3}) appears as a criterion for a consistent choice of a
couple of solutions $K_a$ and $K_{\bar a}$ of (\ref{re}), yielding the
commutation of transfer matrices (hereafter to be defined).

Graphically, (\ref{re}) and (\ref{re3}) are represented as follows:

\setlength{\unitlength}{0.0015cm}
\begin{center}
\begingroup\makeatletter\ifx\SetFigFont\undefined%
\gdef\SetFigFont#1#2#3#4#5{%
  \reset@font\fontsize{#1}{#2pt}%
  \fontfamily{#3}\fontseries{#4}\fontshape{#5}%
  \selectfont}%
\fi\endgroup%
{\renewcommand{\dashlinestretch}{30}
\begin{picture}(4227,3308)(0,-10)
\path(4121,2523)(4215,2616)
\path(4121,2428)(4215,2523)
\path(4121,2240)(4215,2335)
\path(4121,2335)(4215,2428)
\path(4121,2147)(4215,2240)
\path(4121,2053)(4215,2147)
\path(4121,1960)(4215,2053)
\path(4121,1865)(4215,1960)
\path(4121,1772)(4215,1865)
\path(4121,1679)(4215,1772)
\path(4121,1585)(4215,1679)
\path(4121,1490)(4215,1585)
\path(4121,1397)(4215,1490)
\path(4121,1304)(4215,1397)
\path(4121,1115)(4215,1211)
\path(4121,1211)(4215,1304)
\path(4121,1023)(4215,1115)
\path(4121,930)(4215,1023)
\path(4121,836)(4215,930)
\path(4121,741)(4215,836)
\path(4121,648)(4215,741)
\path(4121,554)(4215,648)
\path(4121,461)(4215,554)
\path(4121,368)(4215,461)
\path(1310,2523)(1404,2616)
\path(1310,2428)(1404,2523)
\path(1310,2240)(1404,2335)
\path(1310,2335)(1404,2428)
\path(1310,2147)(1404,2240)
\path(1310,2053)(1404,2147)
\path(1310,1960)(1404,2053)
\path(1310,1865)(1404,1960)
\path(1310,1772)(1404,1865)
\path(1310,1679)(1404,1772)
\path(1310,1585)(1404,1679)
\path(1310,1490)(1404,1585)
\path(1310,1397)(1404,1490)
\path(1310,1304)(1404,1397)
\path(1310,1115)(1404,1211)
\path(1310,1211)(1404,1304)
\path(1310,1023)(1404,1115)
\path(1310,930)(1404,1023)
\path(1310,836)(1404,930)
\path(1310,741)(1404,836)
\path(1310,648)(1404,741)
\path(1310,554)(1404,648)
\path(1310,461)(1404,554)
\path(1310,368)(1404,461)
\path(1310,2616)(1404,2709)
\path(1310,2616)(1404,2709)
\path(1310,2709)(1404,2802)
\path(1310,2709)(1404,2802)
\path(1310,2802)(1404,2898)
\path(1310,2802)(1404,2898)
\path(1310,2898)(1404,2989)
\path(1310,2898)(1404,2989)
\path(4121,3252)(4121,368)
\path(4121,2616)(4215,2709)
\path(4121,2616)(4215,2709)
\path(4121,2709)(4215,2802)
\path(4121,2709)(4215,2802)
\path(4121,2802)(4215,2898)
\path(4121,2802)(4215,2898)
\path(4121,2898)(4215,2989)
\path(4121,2898)(4215,2989)
\path(1881,1772)(2116,1772)
\path(1881,1772)(2116,1772)
\path(1881,1633)(2116,1633)
\path(1881,1633)(2116,1633)
\path(372,45)(1310,1679)(560,2989)
\path(1310,3252)(1310,368)
\path(1310,2999)(1404,3093)
\path(1310,2999)(1404,3093)
\path(1310,3093)(1404,3187)
\path(1310,3093)(1404,3187)
\path(1310,3187)(1404,3281)
\path(1310,3187)(1404,3281)
\path(4121,2999)(4215,3093)
\path(4121,2999)(4215,3093)
\path(4121,3093)(4215,3187)
\path(4121,3093)(4215,3187)
\path(4121,3187)(4215,3281)
\path(4121,3187)(4215,3281)
\path(3372,359)(4121,1672)(3372,2981)
\path(12,315)(1325,876)(57,1440)
\path(2802,1935)(4115,2496)(2850,3058)
\put(237,0){\makebox(0,0)[lb]{\smash{{{\SetFigFont{12}{14.4}{\rmdefault}
{\mddefault}{\updefault}b}}}}}
\put(12,405){\makebox(0,0)[lb]{\smash{{{\SetFigFont{12}{14.4}{\rmdefault}
{\mddefault}{\updefault}a}}}}}
\put(3207,360){\makebox(0,0)[lb]{\smash{{{\SetFigFont{12}{14.4}{\rmdefault}
{\mddefault}{\updefault}b}}}}}
\put(2847,2025){\makebox(0,0)[lb]{\smash{{{\SetFigFont{12}{14.4}{\rmdefault}
{\mddefault}{\updefault}a}}}}}
\end{picture}
}
\qquad
\qquad
{\renewcommand{\dashlinestretch}{30}
\begin{picture}(4594,3287)(0,-10)
\path(4488,2445)(4582,2538)
\path(4488,2350)(4582,2445)
\path(4488,2162)(4582,2257)
\path(4488,2257)(4582,2350)
\path(4488,2069)(4582,2162)
\path(4488,1975)(4582,2069)
\path(4488,1882)(4582,1975)
\path(4488,1787)(4582,1882)
\path(4488,1694)(4582,1787)
\path(4488,1601)(4582,1694)
\path(4488,1507)(4582,1601)
\path(4488,1412)(4582,1507)
\path(4488,1319)(4582,1412)
\path(4488,1226)(4582,1319)
\path(4488,1037)(4582,1133)
\path(4488,1133)(4582,1226)
\path(4488,945)(4582,1037)
\path(4488,852)(4582,945)
\path(4488,758)(4582,852)
\path(4488,663)(4582,758)
\path(4488,570)(4582,663)
\path(4488,476)(4582,570)
\path(4488,383)(4582,476)
\path(4488,290)(4582,383)
\path(1677,2445)(1771,2538)
\path(1677,2350)(1771,2445)
\path(1677,2162)(1771,2257)
\path(1677,2257)(1771,2350)
\path(1677,2069)(1771,2162)
\path(1677,1975)(1771,2069)
\path(1677,1882)(1771,1975)
\path(1677,1787)(1771,1882)
\path(1677,1694)(1771,1787)
\path(1677,1601)(1771,1694)
\path(1677,1507)(1771,1601)
\path(1677,1412)(1771,1507)
\path(1677,1319)(1771,1412)
\path(1677,1226)(1771,1319)
\path(1677,1037)(1771,1133)
\path(1677,1133)(1771,1226)
\path(1677,945)(1771,1037)
\path(1677,852)(1771,945)
\path(1677,758)(1771,852)
\path(1677,663)(1771,758)
\path(1677,570)(1771,663)
\path(1677,476)(1771,570)
\path(1677,383)(1771,476)
\path(1677,290)(1771,383)
\path(1677,2538)(1771,2631)
\path(1677,2538)(1771,2631)
\path(1677,2631)(1771,2724)
\path(1677,2631)(1771,2724)
\path(1677,2724)(1771,2820)
\path(1677,2724)(1771,2820)
\path(1677,2820)(1771,2911)
\path(1677,2820)(1771,2911)
\put(59.747,1466.255){\arc{202.015}{4.1745}{6.0965}}
\put(259.545,1507.945){\arc{206.260}{1.0596}{2.9172}}
\put(359.518,1325.943){\arc{209.059}{4.2189}{6.0411}}
\put(560.225,1372.279){\arc{202.962}{1.0354}{2.9303}}
\put(661.609,1193.508){\arc{208.154}{4.2155}{6.0555}}
\put(863.021,1239.517){\arc{205.048}{1.0616}{2.9202}}
\put(962.083,1056.963){\arc{210.381}{4.2269}{6.0331}}
\put(1160.991,1099.979){\arc{196.932}{1.0023}{2.9683}}
\put(1465.984,971.987){\arc{206.700}{1.0767}{2.9272}}
\put(1263.518,924.943){\arc{209.059}{4.2189}{6.0411}}
\put(1569.518,789.943){\arc{209.059}{4.2189}{6.0411}}
\put(1569.518,789.943){\arc{209.059}{4.2189}{6.0411}}
\put(2870.747,1525.745){\arc{202.014}{0.1867}{2.1087}}
\put(3070.545,1484.054){\arc{206.262}{3.3660}{5.2236}}
\put(3170.518,1666.057){\arc{209.060}{0.2421}{2.0643}}
\put(3371.225,1619.721){\arc{202.961}{3.3528}{5.2478}}
\put(3472.609,1798.492){\arc{208.152}{0.2277}{2.0676}}
\put(3674.021,1752.483){\arc{205.048}{3.3630}{5.2216}}
\put(3773.083,1935.037){\arc{210.381}{0.2501}{2.0563}}
\put(3971.991,1892.021){\arc{196.931}{3.3149}{5.2809}}
\put(59.747,207.745){\arc{202.014}{0.1867}{2.1087}}
\put(259.545,166.054){\arc{206.262}{3.3660}{5.2236}}
\put(359.518,348.057){\arc{209.060}{0.2421}{2.0643}}
\put(560.224,301.721){\arc{202.962}{3.3528}{5.2478}}
\put(661.609,480.492){\arc{208.152}{0.2277}{2.0676}}
\put(863.021,434.483){\arc{205.048}{3.3630}{5.2216}}
\put(962.083,617.037){\arc{210.381}{0.2501}{2.0563}}
\put(1160.991,574.021){\arc{196.931}{3.3149}{5.2809}}
\put(2874.747,2791.255){\arc{202.015}{4.1745}{6.0965}}
\put(3074.545,2832.945){\arc{206.260}{1.0596}{2.9172}}
\put(3174.518,2650.943){\arc{209.059}{4.2189}{6.0411}}
\put(3375.225,2697.279){\arc{202.962}{1.0354}{2.9303}}
\put(3476.609,2518.508){\arc{208.154}{4.2155}{6.0555}}
\put(3678.021,2564.517){\arc{205.048}{1.0616}{2.9202}}
\put(3777.083,2381.963){\arc{210.381}{4.2269}{6.0331}}
\put(3975.991,2424.979){\arc{196.932}{1.0023}{2.9683}}
\put(4280.984,2296.987){\arc{206.700}{1.0767}{2.9272}}
\put(4078.518,2249.943){\arc{209.059}{4.2189}{6.0411}}
\put(4384.518,2114.943){\arc{209.059}{4.2189}{6.0411}}
\put(4384.518,2114.943){\arc{209.059}{4.2189}{6.0411}}
\put(4276.984,2020.013){\arc{206.700}{3.3560}{5.2065}}
\put(4074.518,2067.057){\arc{209.060}{0.2421}{2.0643}}
\put(4380.518,2202.057){\arc{209.060}{0.2421}{2.0643}}
\put(4380.518,2202.057){\arc{209.060}{0.2421}{2.0643}}
\put(1465.984,702.013){\arc{206.700}{3.3560}{5.2065}}
\put(1263.518,749.057){\arc{209.060}{0.2421}{2.0643}}
\put(1569.518,884.057){\arc{209.060}{0.2421}{2.0643}}
\put(1569.518,884.057){\arc{209.060}{0.2421}{2.0643}}
\path(4488,3174)(4488,290)
\path(4488,2538)(4582,2631)
\path(4488,2538)(4582,2631)
\path(4488,2631)(4582,2724)
\path(4488,2631)(4582,2724)
\path(4488,2724)(4582,2820)
\path(4488,2724)(4582,2820)
\path(4488,2820)(4582,2911)
\path(4488,2820)(4582,2911)
\path(2248,1694)(2483,1694)
\path(2248,1694)(2483,1694)
\path(2248,1555)(2483,1555)
\path(2248,1555)(2483,1555)
\path(784,12)(1677,1601)(927,2911)
\path(1677,2921)(1771,3015)
\path(1677,2921)(1771,3015)
\path(1677,3015)(1771,3109)
\path(1677,3015)(1771,3109)
\path(1677,3109)(1771,3203)
\path(1677,3109)(1771,3203)
\path(4488,2921)(4582,3015)
\path(4488,2921)(4582,3015)
\path(4488,3015)(4582,3109)
\path(4488,3015)(4582,3109)
\path(4488,3109)(4582,3203)
\path(4488,3109)(4582,3203)
\path(3739,281)(4488,1594)(3739,2903)
\path(1668,3260)(1676,230)
\put(19,237){\makebox(0,0)[lb]{\smash{{{\SetFigFont{12}{14.4}{\rmdefault}
{\mddefault}{\updefault}a}}}}}
\put(1054,192){\makebox(0,0)[lb]{\smash{{{\SetFigFont{12}{14.4}{\rmdefault}
{\mddefault}{\updefault}b}}}}}
\put(3619,372){\makebox(0,0)[lb]{\smash{{{\SetFigFont{12}{14.4}{\rmdefault}
{\mddefault}{\updefault}b}}}}}
\put(2854,1632){\makebox(0,0)[lb]{\smash{{{\SetFigFont{12}{14.4}{\rmdefault}
{\mddefault}{\updefault}a}}}}}
\end{picture}
}
\end{center}
These $K$ matrices (solutions of the soliton preserving reflection equation
(\ref{re})) have been classified for $sl(\enne)$ Yangians in \cite{Mint}.
This classification can be recovered as a particular case of our
proposition \ref{prop:SP1}, where the $sl(\emme|\enne)$ Yangians are
studied. Yang--Baxter and reflection equations will indeed take the same
form albeit with a graded tensor product in the superalgebraic case (see
section \ref{sect:slmn} for more details). Proposition \ref{prop:SP2} then
provides the classification of pairs $\{K_a(\lambda),K_{\bar a}(\lambda)\}$
which obey (\ref{re}) and the compatibility equation (\ref{re3}).

\vspace*{3mm}

\subsubsection{Soliton non-preserving reflection matrices}

In the context of soliton non-preserving boundary conditions one has to
consider \cite{boco, gand, doikou1} the case where a soliton reflects back
as an anti-soliton. The corresponding reflection equation has the form
\begin{equation}
  R_{ab}(\lambda_{a}-\lambda_{b})\ \tK_{a}(\lambda_{a})\ \bar
  R_{ba}(\lambda_{a}+\lambda_{b})\ \tK_{b}(\lambda_{b}) =
  \tK_{b}(\lambda_{b})\ \bar R_{ab}(\lambda_{a}+\lambda_{b})\
  \tK_{a}(\lambda_{a})\ R_{ba}(\lambda_{a}-\lambda_{b}).
\label{re2}
\end{equation}
Note that equation (\ref{re2}) is satisfied by the generators of the so-called
\textit{twisted Yangian} \cite{Ytwist,MNO}, which will be discussed in
section \ref{sect:symTw}.

Similarly to the previous case, one introduces $\tK_{\bar a}$, describing an
anti-soliton that reflects back as a soliton, satisfying (\ref{re2}) and the
consistency condition
\begin{equation}
  \bar R_{ab}(\lambda_{a}-\lambda_{b})\ \tK_{\bar a}(\lambda_{a})\
  R_{ba}(\lambda_{a}+\lambda_{b})\ \tK_{b}(\lambda_{b}) =
  \tK_{b}(\lambda_{b})\ R_{ab}(\lambda_{a}+\lambda_{b})\ \tK_{\bar
  a}(\lambda_{a})\ \bar R_{ba}(\lambda_{a}-\lambda_{b}) \;.
\label{re4}
\end{equation}
Graphically, (\ref{re2}) and (\ref{re4}) are represented as follows:

%%\begin{figure}[htbp]
%%  \centering
\setlength{\unitlength}{0.0015cm}
\begin{center}
{\renewcommand{\dashlinestretch}{30}
\begin{picture}(4604,3335)(0,-10)
\path(4498,2493)(4592,2586)
\path(4498,2398)(4592,2493)
\path(4498,2210)(4592,2305)
\path(4498,2305)(4592,2398)
\path(4498,2117)(4592,2210)
\path(4498,2023)(4592,2117)
\path(4498,1930)(4592,2023)
\path(4498,1835)(4592,1930)
\path(4498,1742)(4592,1835)
\path(4498,1649)(4592,1742)
\path(4498,1555)(4592,1649)
\path(4498,1460)(4592,1555)
\path(4498,1367)(4592,1460)
\path(4498,1274)(4592,1367)
\path(4498,1085)(4592,1181)
\path(4498,1181)(4592,1274)
\path(4498,993)(4592,1085)
\path(4498,900)(4592,993)
\path(4498,806)(4592,900)
\path(4498,711)(4592,806)
\path(4498,618)(4592,711)
\path(4498,524)(4592,618)
\path(4498,431)(4592,524)
\path(4498,338)(4592,431)
\path(1687,2493)(1781,2586)
\path(1687,2398)(1781,2493)
\path(1687,2210)(1781,2305)
\path(1687,2305)(1781,2398)
\path(1687,2117)(1781,2210)
\path(1687,2023)(1781,2117)
\path(1687,1930)(1781,2023)
\path(1687,1835)(1781,1930)
\path(1687,1742)(1781,1835)
\path(1687,1649)(1781,1742)
\path(1687,1555)(1781,1649)
\path(1687,1460)(1781,1555)
\path(1687,1367)(1781,1460)
\path(1687,1274)(1781,1367)
\path(1687,1085)(1781,1181)
\path(1687,1181)(1781,1274)
\path(1687,993)(1781,1085)
\path(1687,900)(1781,993)
\path(1687,806)(1781,900)
\path(1687,711)(1781,806)
\path(1687,618)(1781,711)
\path(1687,524)(1781,618)
\path(1687,431)(1781,524)
\path(1687,338)(1781,431)
\path(1687,2586)(1781,2679)
\path(1687,2586)(1781,2679)
\path(1687,2679)(1781,2772)
\path(1687,2679)(1781,2772)
\path(1687,2772)(1781,2868)
\path(1687,2772)(1781,2868)
\path(1687,2868)(1781,2959)
\path(1687,2868)(1781,2959)
\put(59.747,1574.255){\arc{202.015}{4.1745}{6.0965}}
\put(259.545,1615.945){\arc{206.260}{1.0596}{2.9172}}
\put(359.518,1433.943){\arc{209.059}{4.2189}{6.0411}}
\put(560.225,1480.279){\arc{202.962}{1.0354}{2.9303}}
\put(661.609,1301.508){\arc{208.154}{4.2155}{6.0555}}
\put(863.021,1347.517){\arc{205.048}{1.0616}{2.9202}}
\put(962.083,1164.963){\arc{210.381}{4.2269}{6.0331}}
\put(1160.991,1207.979){\arc{196.932}{1.0023}{2.9683}}
\put(1465.984,1079.987){\arc{206.700}{1.0767}{2.9272}}
\put(1263.518,1032.943){\arc{209.059}{4.2189}{6.0411}}
\put(1569.518,897.943){\arc{209.059}{4.2189}{6.0411}}
\put(1569.518,897.943){\arc{209.059}{4.2189}{6.0411}}
\put(2884.747,2839.255){\arc{202.015}{4.1745}{6.0965}}
\put(3084.545,2880.945){\arc{206.260}{1.0596}{2.9172}}
\put(3184.518,2698.943){\arc{209.059}{4.2189}{6.0411}}
\put(3385.225,2745.279){\arc{202.962}{1.0354}{2.9303}}
\put(3486.609,2566.508){\arc{208.154}{4.2155}{6.0555}}
\put(3688.021,2612.517){\arc{205.048}{1.0616}{2.9202}}
\put(3787.083,2429.963){\arc{210.381}{4.2269}{6.0331}}
\put(3985.991,2472.979){\arc{196.932}{1.0023}{2.9683}}
\put(4290.984,2344.987){\arc{206.700}{1.0767}{2.9272}}
\put(4088.518,2297.943){\arc{209.059}{4.2189}{6.0411}}
\put(4394.518,2162.943){\arc{209.059}{4.2189}{6.0411}}
\put(4394.518,2162.943){\arc{209.059}{4.2189}{6.0411}}
\put(3663.706,2975.059){\arc{200.940}{4.8150}{6.7371}}
\put(3844.781,2889.510){\arc{199.626}{1.6488}{3.5703}}
\put(3821.358,2685.600){\arc{211.130}{4.8611}{6.6474}}
\put(4013.031,2607.046){\arc{203.293}{1.6795}{3.5563}}
\put(3987.198,2400.319){\arc{213.424}{4.8515}{6.6706}}
\put(4179.036,2319.592){\arc{202.865}{1.6997}{3.5513}}
\put(4152.754,2114.631){\arc{210.413}{4.8386}{6.6490}}
\put(4337.379,2032.856){\arc{194.011}{1.6263}{3.6140}}
\put(4509.477,1748.967){\arc{203.998}{1.7132}{3.5771}}
\put(4315.012,1828.886){\arc{216.907}{4.8697}{6.6302}}
\put(848.706,2990.059){\arc{200.940}{4.8150}{6.7371}}
\put(1029.781,2904.510){\arc{199.626}{1.6488}{3.5703}}
\put(1006.358,2700.600){\arc{211.130}{4.8611}{6.6474}}
\put(1198.031,2622.046){\arc{203.293}{1.6795}{3.5563}}
\put(1172.198,2415.319){\arc{213.424}{4.8515}{6.6706}}
\put(1364.036,2334.592){\arc{202.865}{1.6997}{3.5513}}
\put(1337.754,2129.631){\arc{210.413}{4.8386}{6.6490}}
\put(1522.379,2047.856){\arc{194.011}{1.6263}{3.6140}}
\put(1694.477,1763.967){\arc{203.998}{1.7132}{3.5771}}
\put(1500.012,1843.886){\arc{216.907}{4.8697}{6.6302}}
\path(4498,3222)(4498,338)
\path(4498,2586)(4592,2679)
\path(4498,2586)(4592,2679)
\path(4498,2679)(4592,2772)
\path(4498,2679)(4592,2772)
\path(4498,2772)(4592,2868)
\path(4498,2772)(4592,2868)
\path(4498,2868)(4592,2959)
\path(4498,2868)(4592,2959)
\path(2258,1742)(2493,1742)
\path(2258,1742)(2493,1742)
\path(2258,1603)(2493,1603)
\path(2258,1603)(2493,1603)
\path(1687,2969)(1781,3063)
\path(1687,2969)(1781,3063)
\path(1687,3063)(1781,3157)
\path(1687,3063)(1781,3157)
\path(1687,3157)(1781,3251)
\path(1687,3157)(1781,3251)
\path(4498,2969)(4592,3063)
\path(4498,2969)(4592,3063)
\path(4498,3063)(4592,3157)
\path(4498,3063)(4592,3157)
\path(4498,3157)(4592,3251)
\path(4498,3157)(4592,3251)
\path(1678,3308)(1686,278)
\path(29,150)(1694,915)
\path(2826,1470)(4491,2213)
\path(799,60)(1684,1645)
\path(3674,180)(4489,1635)
\put(29,285){\makebox(0,0)[lb]{\smash{{{\SetFigFont{12}{14.4}{\rmdefault}
{\mddefault}{\updefault}a}}}}}
\put(2819,1570){\makebox(0,0)[lb]{\smash{{{\SetFigFont{12}{14.4}{\rmdefault}
{\mddefault}{\updefault}a}}}}}
\put(944,0){\makebox(0,0)[lb]{\smash{{{\SetFigFont{12}{14.4}{\rmdefault}
{\mddefault}{\updefault}b}}}}}
\put(3509,300){\makebox(0,0)[lb]{\smash{{{\SetFigFont{12}{14.4}{\rmdefault}
{\mddefault}{\updefault}b}}}}}
\end{picture}
}
\qquad
%%\mbox{and}
\qquad
{\renewcommand{\dashlinestretch}{30}
\begin{picture}(4609,3335)(0,-10)
\path(4503,2493)(4597,2586)
\path(4503,2398)(4597,2493)
\path(4503,2210)(4597,2305)
\path(4503,2305)(4597,2398)
\path(4503,2117)(4597,2210)
\path(4503,2023)(4597,2117)
\path(4503,1930)(4597,2023)
\path(4503,1835)(4597,1930)
\path(4503,1742)(4597,1835)
\path(4503,1649)(4597,1742)
\path(4503,1555)(4597,1649)
\path(4503,1460)(4597,1555)
\path(4503,1367)(4597,1460)
\path(4503,1274)(4597,1367)
\path(4503,1085)(4597,1181)
\path(4503,1181)(4597,1274)
\path(4503,993)(4597,1085)
\path(4503,900)(4597,993)
\path(4503,806)(4597,900)
\path(4503,711)(4597,806)
\path(4503,618)(4597,711)
\path(4503,524)(4597,618)
\path(4503,431)(4597,524)
\path(4503,338)(4597,431)
\path(1692,2493)(1786,2586)
\path(1692,2398)(1786,2493)
\path(1692,2210)(1786,2305)
\path(1692,2305)(1786,2398)
\path(1692,2117)(1786,2210)
\path(1692,2023)(1786,2117)
\path(1692,1930)(1786,2023)
\path(1692,1835)(1786,1930)
\path(1692,1742)(1786,1835)
\path(1692,1649)(1786,1742)
\path(1692,1555)(1786,1649)
\path(1692,1460)(1786,1555)
\path(1692,1367)(1786,1460)
\path(1692,1274)(1786,1367)
\path(1692,1085)(1786,1181)
\path(1692,1181)(1786,1274)
\path(1692,993)(1786,1085)
\path(1692,900)(1786,993)
\path(1692,806)(1786,900)
\path(1692,711)(1786,806)
\path(1692,618)(1786,711)
\path(1692,524)(1786,618)
\path(1692,431)(1786,524)
\path(1692,338)(1786,431)
\path(1692,2586)(1786,2679)
\path(1692,2586)(1786,2679)
\path(1692,2679)(1786,2772)
\path(1692,2679)(1786,2772)
\path(1692,2772)(1786,2868)
\path(1692,2772)(1786,2868)
\path(1692,2868)(1786,2959)
\path(1692,2868)(1786,2959)
\put(59.747,218.745){\arc{202.014}{0.1867}{2.1087}}
\put(259.545,177.055){\arc{206.260}{3.3660}{5.2236}}
\put(359.518,359.057){\arc{209.060}{0.2421}{2.0643}}
\put(560.225,312.721){\arc{202.961}{3.3528}{5.2478}}
\put(661.609,491.492){\arc{208.152}{0.2277}{2.0676}}
\put(863.021,445.483){\arc{205.048}{3.3630}{5.2216}}
\put(962.083,628.037){\arc{210.381}{0.2501}{2.0563}}
\put(1160.991,585.021){\arc{196.931}{3.3149}{5.2809}}
\put(1465.984,713.013){\arc{206.700}{3.3560}{5.2065}}
\put(1263.518,760.057){\arc{209.060}{0.2421}{2.0643}}
\put(1569.518,895.057){\arc{209.060}{0.2421}{2.0643}}
\put(1569.518,895.057){\arc{209.060}{0.2421}{2.0643}}
\put(2889.747,1510.745){\arc{202.014}{0.1867}{2.1087}}
\put(3089.545,1469.055){\arc{206.260}{3.3660}{5.2236}}
\put(3189.518,1651.057){\arc{209.060}{0.2421}{2.0643}}
\put(3390.225,1604.721){\arc{202.961}{3.3528}{5.2478}}
\put(3491.609,1783.492){\arc{208.152}{0.2277}{2.0676}}
\put(3693.021,1737.483){\arc{205.048}{3.3630}{5.2216}}
\put(3792.083,1920.037){\arc{210.381}{0.2501}{2.0563}}
\put(3990.991,1877.021){\arc{196.931}{3.3149}{5.2809}}
\put(4295.984,2005.013){\arc{206.700}{3.3560}{5.2065}}
\put(4093.518,2052.057){\arc{209.060}{0.2421}{2.0643}}
\put(4399.518,2187.057){\arc{209.060}{0.2421}{2.0643}}
\put(4399.518,2187.057){\arc{209.060}{0.2421}{2.0643}}
\put(848.706,2975.059){\arc{200.940}{4.8150}{6.7371}}
\put(1029.781,2889.510){\arc{199.626}{1.6488}{3.5703}}
\put(1006.358,2685.600){\arc{211.130}{4.8611}{6.6474}}
\put(1198.031,2607.046){\arc{203.293}{1.6795}{3.5563}}
\put(1172.198,2400.319){\arc{213.424}{4.8515}{6.6706}}
\put(1364.036,2319.592){\arc{202.865}{1.6997}{3.5513}}
\put(1337.754,2114.631){\arc{210.413}{4.8386}{6.6490}}
\put(1522.379,2032.856){\arc{194.011}{1.6263}{3.6140}}
\put(1694.477,1748.967){\arc{203.998}{1.7132}{3.5771}}
\put(1500.012,1828.886){\arc{216.907}{4.8697}{6.6302}}
\put(3660.706,2878.059){\arc{200.940}{4.8150}{6.7371}}
\put(3841.781,2792.510){\arc{199.626}{1.6488}{3.5703}}
\put(3818.358,2588.600){\arc{211.130}{4.8611}{6.6474}}
\put(4010.031,2510.046){\arc{203.293}{1.6795}{3.5563}}
\put(3984.198,2303.319){\arc{213.424}{4.8515}{6.6706}}
\put(4176.036,2222.592){\arc{202.865}{1.6997}{3.5513}}
\put(4149.754,2017.631){\arc{210.413}{4.8386}{6.6490}}
\put(4334.379,1935.856){\arc{194.011}{1.6263}{3.6140}}
\put(4506.477,1651.967){\arc{203.998}{1.7132}{3.5771}}
\put(4312.012,1731.886){\arc{216.907}{4.8697}{6.6302}}
\path(4503,2586)(4597,2679)
\path(4503,2586)(4597,2679)
\path(4503,2679)(4597,2772)
\path(4503,2679)(4597,2772)
\path(4503,2772)(4597,2868)
\path(4503,2772)(4597,2868)
\path(4503,2868)(4597,2959)
\path(4503,2868)(4597,2959)
\path(2263,1742)(2498,1742)
\path(2263,1742)(2498,1742)
\path(2263,1603)(2498,1603)
\path(2263,1603)(2498,1603)
\path(1692,2969)(1786,3063)
\path(1692,2969)(1786,3063)
\path(1692,3063)(1786,3157)
\path(1692,3063)(1786,3157)
\path(1692,3157)(1786,3251)
\path(1692,3157)(1786,3251)
\path(4503,2969)(4597,3063)
\path(4503,2969)(4597,3063)
\path(4503,3063)(4597,3157)
\path(4503,3063)(4597,3157)
\path(4503,3157)(4597,3251)
\path(4503,3157)(4597,3251)
\path(1683,3308)(1691,278)
\path(17,1632)(1682,867)
\path(4503,3222)(4503,338)
\path(2831,2890)(4496,2147)
\path(799,75)(1684,1660)
\path(3679,105)(4494,1560)
\put(34,285){\makebox(0,0)[lb]{\smash{{{\SetFigFont{12}{14.4}{\rmdefault}
{\mddefault}{\updefault}a}}}}}
\put(2824,1570){\makebox(0,0)[lb]{\smash{{{\SetFigFont{12}{14.4}{\rmdefault}
{\mddefault}{\updefault}a}}}}}
\put(949,0){\makebox(0,0)[lb]{\smash{{{\SetFigFont{12}{14.4}{\rmdefault}
{\mddefault}{\updefault}b}}}}}
\put(3514,300){\makebox(0,0)[lb]{\smash{{{\SetFigFont{12}{14.4}{\rmdefault}
{\mddefault}{\updefault}b}}}}}
\end{picture}
}
\end{center}
The $K$ matrices corresponding to the soliton non-preserving reflection
equation are classified in proposition \ref{prop:SNP1}, in the case of
$sl(\emme|\enne)$ Yangians, where once again similar Yang--Baxter and
reflection equations occur. Proposition \ref{prop:SNP2} then provides the
classification of pairs $\{\tK_a(\lambda),\tK_{\bar a}(\lambda)\}$ which
obey (\ref{re2}) and the compatibility equation (\ref{re4}).

\section{The transfer matrix\label{sect:transfer}}

We are now in a position to build open spin chains with different boundary
conditions from the objects $K$, $\tK$, $R$ and $\bar R$ \cite{sklyanin}.
Our purpose is to determine the spectrum and the symmetries of the transfer
matrix for the case where soliton non-preserving boundary conditions are
implemented. We first recall the general settings for the soliton
preserving case.

\subsection{Soliton preserving case}

Let us first define the transfer matrix for the well-known boundary
conditions, i.e. the soliton preserving ones. The starting point is the
construction of the monodromy matrices
\begin{eqnarray}
  \cT_a(\lambda) &=& T_{a}(\lambda)\
  K_{a}^{-}(\lambda)\ \hat T_{a}(\lambda), 
  \label{monoSP}\\
  \cTbar_a(\lambda) &=& T_{\bar a}(\lambda)\
  K_{\bar a}^{-}(\lambda)\ \hat T_{\bar a}(\lambda) 
  \label{monoSPbar}
\end{eqnarray}
and two transfer matrices (soliton--soliton and 
anti-soliton--anti-soliton) 
\begin{equation}
  t(\lambda) = \tr_{a} K_{a}^{+}(\lambda)\ \cT_{a}(\lambda), \qquad
  \bar t(\lambda) = \tr_{a} K_{\bar a}^{+}(\lambda)\ \cTbar_{a}(\lambda) \;,
  \label{t1}
\end{equation}
with
\begin{alignat}{4}
  &T_{a}(\lambda) &=& R_{aL}(\lambda) \ldots R_{a1}(\lambda) \;,
  \qquad\qquad && \hat T_{a}(\lambda) &=& R_{1a}(\lambda) \ldots
  R_{La}(\lambda) \;, \nonumber \\
  &T_{\bar a}(\lambda) &=& \bar R_{aL}(\lambda) \ldots \bar R_{a1}(\lambda)
  \;, && \hat T_{\bar a}(\lambda) &=& \bar R_{1a}(\lambda) \ldots \bar
  R_{La}(\lambda)\;.
  \label{T1}
\end{alignat}
The numerical matrices $K_{a}^{-}(\lambda)$, $K_{\bar a}^{-}(\lambda)$ are
solutions of (\ref{re}), (\ref{re3}) and $K_{a}^{+}$ satisfies a reflection
equation `dual' to (\ref{re}),
\begin{eqnarray}
  && R_{ab}(-\lambda_{a}+\lambda_{b})\ K_{a}^{+}(\lambda_{a})^t \
  R_{ba}(-\lambda_{a}-\lambda_{b}-2i\rho)\ K_{b}^{+}(\lambda_{b})^t
  \qquad\qquad 
  \nonumber \\
  &&\qquad\qquad=\ K_{b}^{+}(\lambda_{b})^t \
  R_{ab}(-\lambda_{a}-\lambda_{b}-2i\rho)\ 
  K_{a}^{+}(\lambda_{a})^t \ R_{ba}(-\lambda_{a}+\lambda_{b}) \;.
  \label{re+}
\end{eqnarray}
The solutions of (\ref{re+}) take the form $K_{a}^{+}(\lambda)
=K^t_{a}(-\lambda -i\rho)$, where $K_{a}(\lambda)$ is a solution of
(\ref{re}). Similarly $K_{\bar a}^{+}(\lambda)$ satisfies a reflection
equation dual to (\ref{re}), the solutions of which being of the form
$K_{\bar a}^{+}(\lambda) = K^t_{\bar a}(-\lambda -i\rho) $. In addition,
$K_{\bar a}^{+}$ satisfies also a compatibility condition dual to
(\ref{re3}). Actually, the dual reflection equations happen to be the usual
reflection equations after a redefinition $\lambda_{c} \to -\lambda_{c}
-i\rho$. \\
{}From their explicit expression, one can deduce that the monodromy matrices
$\cT(\lambda)$ and $\cTbar(\lambda)$ obey the following equations:
\begin{eqnarray}
  R_{ab}(\lambda_{a}-\lambda_{b})\ \cT_{a}(\lambda_{a})\
  R_{ba}(\lambda_{a}+\lambda_{b})\ \cT_{b}(\lambda_{b}) &=&
  \cT_{b}(\lambda_{b})\
  R_{ab}(\lambda_{a}+\lambda_{b})\ \cT_{a}(\lambda_{a})\
  R_{ba}(\lambda_{a}-\lambda_{b})\qquad
\label{exchg-monoSP}\\
  R_{ab}(\lambda_{a}-\lambda_{b})\ \cTbar_{a}(\lambda_{a})\
  R_{ba}(\lambda_{a}+\lambda_{b})\ \cTbar_{b}(\lambda_{b}) &=&
  \cTbar_{b}(\lambda_{b})\
  R_{ab}(\lambda_{a}+\lambda_{b})\ \cTbar_{a}(\lambda_{a})\
  R_{ba}(\lambda_{a}-\lambda_{b})\qquad\\
  \bar R_{ab}(\lambda_{a}-\lambda_{b})\ \cTbar_{a}(\lambda_{a})\ \bar
  R_{ba}(\lambda_{a}+\lambda_{b})\ \cT_{b}(\lambda_{b})&=& 
\cT_{b}(\lambda_{b})\
  \bar{R}_{ab}(\lambda_{a}+\lambda_{b})\ 
\cTbar_{a}(\lambda_{a})\ \bar
  R_{ba}(\lambda_{a}-\lambda_{b}) \;,\qquad
\end{eqnarray}
which just correspond to the soliton preserving reflection equations
(\ref{re}) and (\ref{re3}).
One can recognise, in the relation (\ref{exchg-monoSP}), the exchange 
relation of reflection algebras based on $R$-matrix of $\cY(sl_{\enne})$. 
The form of $K^\pm$ determines the precise algebraic structure which 
is involved, see section \ref{sect:sym}.\\

As usual in the framework of spin chain models,
 the commutativity of the transfer matrices (\ref{t1})
\begin{eqnarray}
&[t(\lambda),t(\mu)] = 0&\;\\
{[\bar t(\lambda),\bar t(\mu)]}=0 &\qmbox{}&
[t(\lambda),\bar t(\mu)]=0 \label{t-tbar}
\end{eqnarray}
is ensured by the above exchange relations. The first commutator
guarantees the integrability of the model, 
whose Hamiltonian is given by
\begin{eqnarray}
  {\cal H} = -{1\over 2} {d\over d \lambda}t(\lambda)\Big\vert_{\lambda =0}\;,
\end{eqnarray}
the locality being ensured because $R(0) =\cP$. \\
The commutators (\ref{t-tbar}) will be needed so that the fusion procedure 
be well-defined (see appendices).
\\
The transfer matrix for $K^+(u)=K^-(u)=1$ 
satisfies a crossing-like relation (see for details
\cite{mnanal, doikou1})
\begin{eqnarray}
  t(\lambda) = \bar t(-\lambda -i\rho) \;.
  \label{cross1}
\end{eqnarray}
The eigenvalues of the transfer matrices as well as the corresponding Bethe
Ansatz equations have been derived for diagonal $K$ matrices in \cite{dvgr,
doikou2}. 

\subsection{Soliton non-preserving case}

This case was studied in \cite{doikou1} for the $sl(3)$ chain
only. Here we generalise the results for any $sl(\enne)$. One
introduces two monodromy matrices
\begin{eqnarray}
  \cT_{a}(\lambda) &=& T_{a}(\lambda)\
  \tK_{a}^{-}(\lambda)\ \hat T_{\bar a}(\lambda) \;, \nonumber\\
  \cTbar_a(\lambda) &=& T_{\bar
  a}(\lambda)\ \tK_{\bar a}^{-}(\lambda)\ \hat T_{a}(\lambda) \;,
\end{eqnarray}
and two transfer matrices (anti-soliton--soliton and soliton--anti-soliton)
defined by
\begin{eqnarray}
  t(\lambda) = \tr_{a} \tK_{a}^{+}(\lambda)\ \cT_a(\lambda) \;,\qquad
  \bar t(\lambda) = \tr_{a} \tK_{\bar a}^{+}(\lambda)\ \cTbar_{a}(\lambda) \;,
  \label{t2}
\end{eqnarray}
where now
\begin{alignat}{4}
  &T_{a}(\lambda) &=& R_{a\,2L}(\lambda) \bar R_{a\,2L-1}(\lambda) \ldots
  R_{a\,2}(\lambda) \bar R_{a\,1}(\lambda) \;, \qquad
  &&
  \hat T_{\bar a}(\lambda) &=& R_{1\,a}(\lambda)\bar R_{2\,a}(\lambda) \ldots
  R_{2L-1\,a}(\lambda) \bar R_{2L\,a}(\lambda) \;, \nonumber \\
  &T_{\bar a}(\lambda) &=& \bar R_{a\,2L}(\lambda) R_{a\,2L-1}(\lambda) \ldots
  \bar R_{a2}(\lambda) R_{a1}(\lambda) \;, 
  &&
  \hat T_{a}(\lambda) &=& \bar R_{1a}(\lambda)R_{2a}(\lambda) \ldots \bar
  R_{2L-1\,a}(\lambda) R_{2L\,a}(\lambda) \;.
  \label{T2}
\end{alignat}
Note that, in this case, the number of sites is $2L$ because we want to
build an alternating spin chain, which is going to ensure that the
Hamiltonian of the model is local. This construction is similar 
to the one 
introduced in \cite{APSS}, where however a different notion of $\bar
R$ was used.
 \\
The numerical matrices $\tK_{a}^{-}$, $\tK_{\bar a}^{-}$ are solutions of
(\ref{re2}), (\ref{re4}). The numerical matrices $\tK_{a}^{+}$ and
$\tK_{\bar a}^{+}$ are solutions of the following reflection equations:
\begin{eqnarray}
&&  R_{ab}(-\lambda_{a}+\lambda_{b})\ \tK^+_{a}(\lambda_{a})^t\ \bar
  R_{ba}(-\lambda_{a}-\lambda_{b}-2i\rho)\ 
  \tK^+_{b}(\lambda_{b})^t \nonumber\\
&&\qquad\qquad\qquad\qquad  = \tK^+_{b}(\lambda_{b})^t\ \bar
  R_{ab}(-\lambda_{a}-\lambda_{b}-2i\rho)\ 
  \tK^+_{a}(\lambda_{a})^t\ R_{ba}(-\lambda_{a}+\lambda_{b})\,\qquad \\
&&  R_{ab}(-\lambda_{a}+\lambda_{b})\ \tK^+_{\bar a}(\lambda_{a})^t\ \bar
  R_{ba}(-\lambda_{a}-\lambda_{b}-2i\rho)\ 
  \tK^+_{\bar b}(\lambda_{b})^t \nonumber\\
&&\qquad\qquad\qquad\qquad  = \tK^+_{\bar b}(\lambda_{b})^t\ \bar
  R_{ab}(-\lambda_{a}-\lambda_{b}-2i\rho)\ 
  \tK^+_{\bar a}(\lambda_{a})^t\ R_{ba}(-\lambda_{a}+\lambda_{b})\,\qquad \\
&&  \bar R_{ab}(-\lambda_{a}+\lambda_{b})\ \tK^+_{\bar a}(\lambda_{a})^t\ 
  R_{ba}(-\lambda_{a}-\lambda_{b}-2i\rho)\ 
  \tK^+_{b}(\lambda_{b})^t \nonumber\\
&&\qquad\qquad\qquad\qquad  = \tK^+_{b}(\lambda_{b})^t\
  R_{ab}(-\lambda_{a}-\lambda_{b}-2i\rho)\ 
  \tK^+_{\bar a}(\lambda_{a})^t\ \bar R_{ba}(-\lambda_{a}+\lambda_{b})\,\qquad 
\end{eqnarray}
The commutators
\begin{equation}
  [t(\lambda),t(\mu)] = 0\qmbox{,} {[\bar t(\lambda),\bar t(\mu)]}=0
  \qmbox{and} [t(\lambda),\bar t(\mu)]=0 
  \label{truc}
\end{equation}
are ensured by the above exchange relations for $\tK^+$ and the
relations for the monodromy matrices, namely 
\begin{eqnarray}
  R_{ab}(\lambda_{a}-\lambda_{b})\ \cT_{a}(\lambda_{a})\
  \bar R_{ba}(\lambda_{a}+\lambda_{b})\ \cT_{b}(\lambda_{b}) &=&
  \cT_{b}(\lambda_{b})\
  \bar R_{ab}(\lambda_{a}+\lambda_{b})\ \cT_{a}(\lambda_{a})\
  R_{ba}(\lambda_{a}-\lambda_{b})\;,\qquad 
\label{exchg-monoSNP}\\
  R_{ab}(\lambda_{a}-\lambda_{b})\ \cTbar_{a}(\lambda_{a})\
  \bar R_{ba}(\lambda_{a}+\lambda_{b})\ \cTbar_{b}(\lambda_{b}) &=&
  \cTbar_{b}(\lambda_{b})\
  \bar R_{ab}(\lambda_{a}+\lambda_{b})\ \cTbar_{a}(\lambda_{a})\
  R_{ba}(\lambda_{a}-\lambda_{b})\;,\qquad \\
  \bar R_{ab}(\lambda_{a}-\lambda_{b})\ \cTbar_{a}(\lambda_{a})\ 
  R_{ba}(\lambda_{a}+\lambda_{b})\ \cT_{b}(\lambda_{b})&=& 
\cT_{b}(\lambda_{b})\
  {R}_{ab}(\lambda_{a}+\lambda_{b})\ 
\cTbar_{a}(\lambda_{a})\ \bar
  R_{ba}(\lambda_{a}-\lambda_{b}) \;.\qquad
\end{eqnarray}
The relation (\ref{exchg-monoSNP}) has to be compared with the exchange
relation of twisted Yangians based on $R$-matrix of $\cY(sl_{\enne})$. We
will come back to this point in section \ref{sect:sym}.

In the SNP case, one can show \cite{doikou1} that the transfer matrices
for $\tK^+(u)=\tK^-(u)=1$ 
exhibit a crossing symmetry, namely
\begin{eqnarray}
  t(\lambda) =t(-\lambda -i\rho) \;, \qquad \qquad
  \bar t(\lambda) = \bar t(-\lambda -i\rho)
  \;.
\label{cross2}
\end{eqnarray}
Starting from (\ref{T2}) and (\ref{t2}), the Hamiltonian of the alternating
open spin chain is derived as \cite{doikou1, dewo}
\begin{eqnarray}
  {\cal H} = -{1\over 2} {d\over d \lambda} t(\lambda)\ \bar
  t(\lambda) \Big\vert_{\lambda =0} \;.
  \label{ham}
\end{eqnarray}
The locality is again ensured by $R(0) =\cP$, and integrability is 
guaranteed by (\ref{truc}). 

We can explicitly write the open Hamiltonian in terms of the permutation
operator and the $R(0)$ and $\bar R(0)$ matrices. Let us first introduce
some notations:
\begin{eqnarray}
  \bar R_{ij} = \bar R_{ij}(0), ~~\bar R'_{ij} = {d\over d\lambda} \bar
  R_{ij}(\lambda) \Big\vert_{\lambda=0}\quad\mbox{and} ~~\check
  R'_{ij} = \cP_{ij}\ 
  {d\over d\lambda}R_{ij}(\lambda) \Big\vert_{\lambda=0}.
\end{eqnarray} 
After some algebraic manipulations, in particular taking into account that
$\tr_{0}\cP_{0i}\ \bar R_{0i} \propto \II$, we obtain the following
expression for the Hamiltonian (\ref{ham}) (for a detailed proof see
\cite{doikou1})
\begin{eqnarray}
  {\cal H} &\propto &\sum_{j=1}^L \bar R'_{2j-1\ 2j}\ \bar R_{2j-1\
  2j} + \sum_{j=1}^{L -1}\bar R_{2j+1\ 2j+2}\ \check R'_{2j\ 2j+2}\
  \bar R_{2j+1\ 2j+2} \nonumber \\ 
  &+& \sum_{j=1}^{L -1}\bar R_{2j+1\ 2j+2}\ \bar R_{2j-1\ 2j}\ \bar
  R'_{2j-1\ 2j+2}\ \bar R_{2j-1\ 2j+2}\ \bar R_{2j-1\ 2j}\ \bar R_{2j+1\
  2j+2} \nonumber \\ 
  &+& \sum_{j=1}^{L -1}\bar R_{2j+1\ 2j+2}\ \bar R_{2j-1\ 2j}\ \bar R_{2j-1\
  2j+2}\ \check R'_{2j-1\ 2j+1}\ \bar R_{2j-1\ 2j+2}\ \bar R_{2j-1\ 2j}\
  \bar R_{2j+1\ 2j+2} \nonumber \\ 
  &+& \tr_{0}\check R'_{0\ 2L}\ \bar R_{2L-1\ 2L}\ \cP_{0\ 2L-1}\ \bar
  R_{0\ 2L-1}\ \bar R_{2L-1\ 2L} + \bar R_{12}\ \check R'_{12}\ \bar R_{12},
  \label{ham2}
\end{eqnarray}
which is indeed local including terms that describe interaction up to
four first neighbours.

It is easily shown, acting on (\ref{T2}) by full transposition, that
$\bar t(\lambda) \propto t^{t_1 ... t_{2L}}(\lambda)$ provided that
$(\tK_a^-)^t \propto \tK_{\bar a}^-$. 
Eigenvectors of $\cH$ in (\ref{ham2}) are determined by sole
evaluation of eigenvectors of $t(\lambda)$.
We shall therefore only need to consider diagonalisation of
$t(\lambda)$ in what follows.

\section{Symmetry of the transfer matrix\label{sect:sym}}

In the two (SP and SNP) boundary cases, the use of
exchange relations for the monodromy matrices allows us to determine the
 symmetry of the transfer
matrix. For simplicity, we fix $K^+(\lambda)$ (or $\tK^+(\lambda)$) to be
$\II$, leaving $K^-(\lambda)$ (or $\tK^-(\lambda)$) free.

\subsection{Soliton preserving boundary conditions}

In this case, the general form 
for $K^-(\lambda)$ \cite{Mint} is conjugated
(through a constant matrix) to the following diagonal matrix:
\begin{equation}
  K^-(\lambda)=i\xi\II+\lambda\EE \qmbox{with}
  \EE = \diag(\underbrace{+1,...,+1}_m,\underbrace{-1,...,-1}_n) \;.
\end{equation}
In the particular case of diagonal solutions, one recovers the scaling
limits of the solutions obtained in the ``quantum'' case in \cite{dvgr2}.
The monodromy matrix then $\cT(\lambda)$ generates a $B(\enne,n)$
reflection algebra as studied in \cite{momo}. Taking the trace in space
$a$ of the relation (\ref{exchg-monoSP}), we obtain:
\begin{equation}
  (\lambda_a^2-\lambda_b^2)\,[\,t(\lambda_a)\,,\,\cT(\lambda_b)\,] \,=\,
  (2i\,\lambda_a-\enne)\,[\,\cT(\lambda_a)\,,\,\cT(\lambda_b)\,]
  \label{comtT}
\end{equation}
{}From the asymptotic behaviour ($\lambda \to \infty$) of the $R$ matrix
\begin{eqnarray}
  R_{0i}(\lambda) = \lambda \Big(\II + {i\over \lambda}\cP_{0i} \Big) \;,
\end{eqnarray}
we deduce 
\begin{equation}
  \cT(\lambda)=\lambda^{2L+1}\left(\EE+\frac{i}{\lambda}
  \Big(\xi+\sum_{j=1}^L\cB_{0j}\Big) +O(\frac{1}{\lambda^{2}}) \right)
\end{equation}
where $\cB_{0j}=\cP_{0j}\,\EE_0+\EE_0\,\cP_{0j}$. We can write
$\cB_{0j}=\sum_{\alpha,\beta=1}^\enne E_{\alpha\beta}\otimes
b_j^{\alpha\beta}$, with $E_{\alpha\beta}$ the elementary matrices acting
in space $0$, and $b_j^{\alpha\beta}$ realizing the $gl(m)\oplus gl(n)$
algebra (in the space $j$).\\
Picking up the coefficient of $\lambda_b^{2L}$ in the relation
(\ref{comtT}), one then concludes that
\begin{equation}
  \left[\,t(\lambda_a)\,,\,\sum_{j=1}^L b^{\alpha\beta}_{j}\,\right]=0\,,  
\end{equation}
i.e. the transfer matrix commutes with the $gl(m)\oplus gl(n)$ algebra.

\subsection{Soliton non-preserving boundary conditions\label{sect:symTw}}

We have already mentioned that the monodromy matrix $\cT(\lambda)$
satisfies one of the defining relations (\ref{exchg-monoSNP}) for the
twisted Yangians $\cY^{\pm}(\enne)$ \cite{Ytwist,MNO}. The general form for
$\tK^-(\lambda)$ is given in proposition \ref{prop:SNP1}: it is constant
and obeys $(\tK^-)^t=\epsilon \tK^-$ with $\epsilon=\pm1$.

We need to investigate the asymptotic behaviour ($\lambda \to \infty$) 
of the $R$ and $\bar R$ matrices given by (\ref{r}) and (\ref{eq:br})
respectively: 
\begin{eqnarray}
  R_{0i}(\lambda) = \lambda \Big(\II + {i\over \lambda}\cP_{0i} 
  \Big) \qmbox{and } 
  \bar R_{0i}(\lambda) = -
  \lambda \Big(\II + {i\over \lambda} {\eppe}_{0i}  
  \Big) \;,
  \label{asy}
\end{eqnarray}
where we have introduced
\begin{equation}
\eppe_{0i} = \rho\,\II-Q_{0i}\;.
\end{equation} 
Accordingly, the monodromy matrices (\ref{T2}) take the following form
\begin{eqnarray}
  T_{0}(\lambda)&=& (-\lambda^2)^{L}
  \Big (\II+ {i\over \lambda}\sum_{i=1}^L 
  \left(\cP_{0,2i} + \eppe_{0,2i-1} \right)
  + O({1\over \lambda^{2}})\Big ), \nonumber \\
  \hat T_{\bar 0}(\lambda) &=& (-\lambda^2)^{L} 
  \Big(\II +{i\over  \lambda}
  \sum_{i=1}^L 
  \left( \eppe_{0,2i} + \cP_{0,2i-1} \right)
  +O({1\over \lambda^{2}})\Big )
  \label{sym1}
\end{eqnarray} 
and finally
\begin{eqnarray}
  T_{0}(\lambda)\ \tK^- \ \hat T_{\bar 0}(\lambda) = 
  \lambda^{4L} \Big (
  \tK^- + {i\over \lambda} 
  \sum_{i=1}^{2L} \cS_{0i}+O({1\over \lambda^{2}}) 
  \Big),
  \label{sym2}
\end{eqnarray}
where 
\begin{equation}
  \label{eq:S02i}
  \cS_{0,2i}=\cP_{0,2i}\tK^- + \tK^- {\eppe_{0,2i}} \;, \qquad
  \cS_{0,2i-1}= \tK^-\cP_{0,2i-1} +  {\eppe_{0,2i-1}}\tK^- \;. 
\end{equation}
Similarly to the previous case, one can show from equation
(\ref{exchg-monoSNP})  that (this time for $\tK^-=1$ only): 
\begin{equation}
  \label{eq:SNPsym}
  \left[ t(\lambda_a),  \sum_{i=1}^{2L} \cS_{0i} \right] = 0
\end{equation}
In this particular case $\cS$ realises the $so(\enne)$ (or $sp(\enne)$)
generators. When $\enne$ is even, as already mentioned, there are two
possibilities for the projector $Q$. More specifically the choice
$\theta_0=1$ in (\ref{eq:V}) corresponds to the $so({\enne})$ case, whereas
$\theta_0=-1$ corresponds to the $sp(\enne)$ case.

\paragraph{Remark:} 
The same construction (open chain with twisted boundary conditions) can be
done starting from the $so(\enne)$, $sp(2\enne)$ and $osp(\emme|2\enne)$
$R$-matrix \cite{soya}. However, since the fundamental representations of
these algebras are self-conjugated, solitons and anti-solitons define 
 the same object. Hence,  the ``twisted''  boundary conditions 
 should be equivalent to open
chains with ``ordinary'' boundary conditions. Indeed, it has been shown in
\cite{soya} that boundary reflection equations (defining the boundary
algebra) and twisted reflection equations (defining the twisted Yangian)
are identical.

\section{Spectrum of the transfer matrix}

Our purpose is to determine the spectrum of the transfer matrix for the
$sl({\enne})$ case.  

\subsection{Treatment of non-diagonal reflection matrices (SP case) \label{sect:diagoK}}
In the soliton preserving case, the classification of reflection matrices 
associated to the Yangian $Y(gl(\enne))$ has been computed in \cite{Mint}. 
It can been recovered from the $sl(\emme|\enne)$ case given in proposition 
\ref{prop:SP1}. Using this classification, it is easy to show the 
following proposition \cite{lepetit}:
\begin{proposition}
Let $K(\lambda)$ be any diagonalizable reflection matrix. $D(\lambda)$,
the corresponding diagonal reflection matrix, can be written as
\begin{equation}
D(\lambda)=U^{-1}\,K(\lambda)\,U
\end{equation}
where $U$, the diagonalization matrix, is constant. 
Let $t_K(\lambda)=\tr_a(T_a(\lambda)K_a(\lambda)\hat{T}_a(\lambda))$ and 
 $t_D(\lambda)=\tr_a(T_a(\lambda)D_a(\lambda)\hat{T}_a(\lambda))$ be the 
corresponding transfer matrices (we set $K_{+}=\II$).\\
Then, $t_K(\lambda)$ and $t_D(\lambda)$ have the same eigenvalues, their 
eigenvectors (say $v_K$ and $v_D$ respectively) being related through
\begin{equation}
v_K = U_1U_2\ldots U_L\,v_D \label{eq:vK-vD}
\end{equation}
\end{proposition}
\textit{Proof}: The fact that the diagonalization matrix is a constant 
(in $\lambda$) is a consequence of the classification (see \cite{Mint} and 
proposition \ref{prop:SP1}). Using the property (\ref{RAA}), one can show that
\begin{equation}
t_K(\lambda)=U_1U_2\ldots U_L\,t_D(\lambda)\,(U_1U_2\ldots U_L)^{-1}\;,
\end{equation}
which is enough to end the proof.
\finproof\\
The general treatment (including the super case) for diagonal reflection 
matrices is done in section \ref{sect:SPK}. From the above property, 
this treatment (for diagonal matrices) is enough to obtain
the  spectrum for \textit{all} the transfer 
matrices associated to \textit{all} the reflection 
matrices (provided they are diagonalisable).\\
As an illustration of this proposition, we compute the eigenvalues 
associated to the non-diagonal reflection matrix \cite{cherednik,abad}
\begin{equation}
%%% c'=2c+1 par rapport a la notation abad-rios
K(\lambda)=\left(\begin{array}{ccccc}
-\lambda+i\xi & 0 & \cdots &  0 & 2k\lambda\\
0 & c\,\lambda+i\xi & \ddots & & 0 \\
\vdots & \ddots &\ddots & \ddots & \vdots \\
0 & & \ddots & c\,\lambda+i\xi & 0\\
2k\lambda & 0 & \cdots &0 & \lambda+i\xi
\end{array}\right)
\qmbox{with} c^2=4k^2+1
\end{equation}
It is easy to see that $K(\lambda)$ is diagonalized by the constant matrix
\begin{equation}
    U=\left(\begin{array}{ccccc}
-\frac{k}{\xi} & 0 & \cdots &  0 & \frac{k}{\xi}\\
0 & 1 & \ddots & & 0 \\
\vdots & \ddots &\ddots & \ddots & \vdots \\
0 & & \ddots & 1& 0\\
\frac{c-1}{2c} & 0 & \cdots &0 & \frac{c+1}{2c}
\end{array}\right)\,. \label{eq:U}
\end{equation}
The corresponding diagonal matrix is given by
\begin{equation}
D(\lambda)=c\,\diag(-\lambda+i\xi',\lambda+i\xi',\ldots,
\lambda+i\xi')\,,\qmbox{with} \xi'=\frac{\xi}{c}\,,
\end{equation}
in accordance with the classification of 
reflection matrices. The $t_{K}(\lambda)$ eigenvalues as well as the Bethe 
equations are identical to the ones of $t_{D}(\lambda)$.
They can be 
deduced from the general treatment given in section \ref{sect:SPK}, 
taking formally $n_{1}=n_{2}=\enne=0$ and specifying $m_{1}=1$. They 
can also be viewed as the scaling limit of quantum groups diagonal solutions \cite{deve}.
The eigenvectors are related using the formula (\ref{eq:vK-vD}), with the explicit 
form (\ref{eq:U}) for $U$.\\[2mm]

This general procedure can be applied for an arbitrary spin chain, 
provided the diagonalization matrix is independent from the spectral 
parameter and commutes, see eq. (\ref{RAA}), with the $R$-matrix under consideration.
When $K_{+}(\lambda)$ is not $\II$, this technics can also be used 
if $K_{+}(\lambda)$ and $K_{-}(\lambda)$ can be
diagonalized in the same basis \cite{gama}. Note however that the 
classification does not ensure the full generality of such an 
assumption.

\subsection{Pseudo-vacuum and dressing functions}
We present below the case when 
soliton non-preserving boundary conditions are
implemented with the simplest choice $\tK^\pm=\II$. 
Results for more general choices of $\tK^-$ can be deduced from the
superalgebra case treated in section \ref{sect:SNPK}. 
\\
We first derive the pseudo-vacuum eigenvalue denoted as
$\Lambda^{0}(\lambda)$, with the pseudo-vacuum being
\begin{eqnarray}
  \vert \omega_{+} \rangle = \bigotimes_{i=1}^{2L} \vert + \rangle _{i}
  \qquad \mbox{where} \qquad \vert + \rangle = \left (
  \begin{array}{c}
      1 \\
      0 \\
      \vdots \\
      0 \\
  \end{array}
  \right)\,\in\,\CC^{\enne}\,. \label{pseudo}
\end{eqnarray}
Note that the pseudo-vacuum is an exact eigenstate of the transfer matrix
(\ref{t1}), and $\Lambda^{0}(\lambda)$ is given by the following expression
\begin{eqnarray}
  \Lambda^{0}(\lambda) = (a(\lambda)\bar b(\lambda))^{2L} g_{0}(\lambda)
  +(b(\lambda)\bar b(\lambda))^{2L}\sum_{l=1}^{{\enne}-2} g_{l}(\lambda)+
  (\bar a(\lambda) b(\lambda))^{2L} g_{{\enne}-1}(\lambda). 
  \label{eigen0}
\end{eqnarray} 
with
\begin{eqnarray}
  a(\lambda) =\lambda+i, ~~b(\lambda)=\lambda, ~~\bar a(\lambda) =
  a(-\lambda - i\rho ), ~~\bar b(\lambda) = b(-\lambda -i\rho) 
  \label{abaBbB}
\end{eqnarray} 
and
\begin{eqnarray}
  g_{l}(\lambda) &=& {\lambda + {i\over 2}(\rho -\theta_0) \over
  \lambda + {i\rho \over 2}}, \qquad 0 \le l < {\enne-1 \over 2} \nonumber \\
  g_{{\enne-1 \over 2}}(\lambda) &=&1,\qquad \qquad \mbox{for $\enne$ odd}
  \nonumber \\ 
  g_{l}(\lambda) &=& g_{\enne-l-1}(-\lambda-i\rho) \;. 
  \label{g}
\end{eqnarray}
We remind that $\rho=\frac{\enne}{2}$.
\\
We make at this point the assumption that any eigenvalue of the transfer
matrix can be written as
\begin{eqnarray}
  \Lambda(\lambda) = (a(\lambda)\bar b(\lambda))^{2L}
  g_{0}(\lambda)A_{0}(\lambda) +(b(\lambda)\bar
  b(\lambda))^{2L}\sum_{l=1}^{{\enne}-2}g_{l}(\lambda)A_{l}(\lambda)+
  (\bar a(\lambda) b(\lambda))^{2L}
  g_{{\enne}-1}(\lambda)A_{{\enne}-1}(\lambda) 
  \qquad
  \label{eigen}
\end{eqnarray}
where the so-called ``dressing functions'' $A_{i}(\lambda)$ need now to be
determined. \\
We immediately get from the crossing symmetry (\ref{cross2}) of the transfer matrix:
\begin{eqnarray}
  A_{0}(\lambda) = A_{{\enne}-1}(-\lambda -i\rho), \qquad A_{l}(\lambda) =
  A_{{\enne}-l-1}(-\lambda -i\rho) \,.
  \label{11}
\end{eqnarray}
Moreover, we obtain from the fusion relation (\ref{ft1}) the following
identity, by a comparison of the forms (\ref{eigen}) for the initial and
fused auxiliary spaces:
\begin{eqnarray}
  A_{0}(\lambda+i\rho)A_{{\enne}-1}(\lambda) =1 \,.
  \label{22}
\end{eqnarray}
Gathering the above two equations (\ref{11}), (\ref{22}) we conclude
\begin{eqnarray}
  A_{0}(\lambda)A_{0}(-\lambda) = 1 \,.
  \label{33}
\end{eqnarray}
Finally from equations (\ref{fusiongen}) important relations between the
dressing functions are entailed for both soliton preserving and soliton
non-preserving boundary conditions. In particular,
\begin{eqnarray}
  \prod_{l=0}^{{\enne}-1} A_{l}(\lambda +i({\enne}-1) -il) =1.\label{conge}
\end{eqnarray}

Taking into account the constraints (\ref{11}), (\ref{33}) and (\ref{conge})
one derives the dressing functions:
\begin{eqnarray}
  A_{0}(\lambda) &=& \prod_{j=1}^{M^{(1)}}{\lambda+
  \lambda_{j}^{(1)}-{i\over 2}\over \lambda+ \lambda_{j}^{(1)} +{i\over 2}}\
  {\lambda-\lambda_{j}^{(1)}-{i\over 2} \over \lambda-\lambda_{j}^{(1)}
  +{i\over 2}} \,, \nonumber \\
  A_{l}(\lambda) &=& \prod_{j=1}^{M^{(l)}}
  {\lambda+\lambda_{j}^{(l)}+{il\over 2}+i \over \lambda+ \lambda_{j}^{(l)}
  +{il\over2}} \; {\lambda-\lambda_{j}^{(l)}+{il\over 2}+i\over \lambda-
  \lambda_{j}^{(l)} +{il\over 2}} \nonumber \\
  && \times \prod_{j=1}^{M^{(l+1)}}{\lambda+ \lambda_{j}^{(l+1)}+{il\over
  2}-{i\over 2}\over \lambda+ \lambda_{j}^{(l+1)} +{il\over 2} +{i\over 2}}\
  {\lambda-\lambda_{j}^{(l+1)}+{il \over 2}-{i\over 2} \over
  \lambda-\lambda_{j}^{(l+1)} + {il\over 2}+{i\over 2}} \,, \qquad 
  1\leq l<\frac{\enne-1}{2} 
\end{eqnarray}
together with the property $A_{l}(\lambda) = A_{\enne-1-l}(-\lambda 
-i\rho)$, and, for $\enne=2n+1$:
\begin{eqnarray}
  A_{n}(\lambda) &=& \prod_{j=1}^{M^{(n)}}
  {\lambda+\lambda_{j}^{(n)}+{in\over 2}+i\over \lambda+ \lambda_{j}^{(n)}
  +{in\over2}} \; {\lambda-\lambda_{j}^{(n)}+{in\over 2}+i\over \lambda-
  \lambda_{j}^{(n)} +{in\over 2}} \ 
  {\lambda+\lambda_{j}^{(n)}+{in\over 2}-{i\over 2} \over \lambda+
  \lambda_{j}^{(n)} +{in\over 2}+{i\over 2}} \;
  {\lambda-\lambda_{j}^{(n)}+{in\over 2}-{i\over 2}\over \lambda-
  \lambda_{j}^{(n)} +{in\over 2}+{i\over 2}}\,,\qquad \label{a20}
\end{eqnarray}
Note that the dressing does not depend on the value of $\theta_{0}$.

The numbers $M^{(l)}$ in (\ref{a20}) are related as
customary to the eigenvalues of diagonal generators $S_{l}$ of the underlying
symmetry algebra (determined in the previous section), namely
\begin{eqnarray}
  S_{1} &=& {1\over 2}M^{(0)}-M^{(1)}, ~~ S_{l} = M^{(l-1)}-M^{(l)}\,\quad
  \mbox{with} \quad S_{l} ={1\over 2}(E_{ll} -E_{\bar l \bar l}),\quad
  1\leq l<\frac{\enne-1}{2}\qquad
  \label{gen}
\end{eqnarray}
with $M^{(0)} =2L $. \\
Recall that for the $sl(\enne)$ case the corresponding numbers $M^{(l)}$,
see e.g. \cite{done}, are given by the following expressions
\begin{eqnarray}
  E_{ll} = M^{(l-1)}-M^{(l)}, \qquad  l=1,\ldots,\enne 
  \label{gen2}
\end{eqnarray}
with $M^{(0)} =2L $ and $M^{(\enne)} =0$. If we now impose $M^{(l)} =
M^{(\enne-l)}$ and consider the differences $E_{ll}-E_{{\bar l} {\bar l}}$, 
we end up with relations (\ref{gen}), in accordance with the folding 
of $sl(\enne)$ leading to $so(\enne)$ and $sp(\enne)$ algebras.

\subsection{Bethe Ansatz equations}

{}From analyticity requirements one obtains the Bethe Ansatz equations
which read as:
\subsubsection{$\bf sl(2n+1)$ algebra}

\begin{eqnarray}
  e_{1}(\lambda_{i}^{(1)})^{2L} &\!\!=\!\!& -\prod_{j=1}^{M^{(1)}}
  e_{2}(\lambda_{i}^{(1)} - \lambda_{j}^{(1)})\ e_{2}(\lambda_{i}^{(1)} +
  \lambda_{j}^{(1)})\ \prod_{ j=1}^{M^{(2)}}e_{-1}(\lambda_{i}^{(1)} -
  \lambda_{j}^{(2)})\ e_{-1}(\lambda_{i}^{(1)} + \lambda_{j}^{(2)})\,,
  \nonumber \\
  1 &\!\!=\!\!& -\prod_{j=1}^{M^{(l)}} e_{2}(\lambda_{i}^{(l)} -
  \lambda_{j}^{(l)})\ e_{2}(\lambda_{i}^{(l)} + \lambda_{j}^{(l)})\ \prod_{
  \tau = \pm 1}\prod_{ j=1}^{M^{(l+\tau)}}e_{-1}(\lambda_{i}^{(l)} -
  \lambda_{j}^{(l+\tau)})\ e_{-1}(\lambda_{i}^{(l)} +
  \lambda_{j}^{(l+\tau)}) \nonumber \\
  && l= 2,\ldots,n-1, \nonumber \\
  e_{-{1\over 2}}(\lambda_{i}^{(n)}) &\!\!=\!\!& -\prod_{j=1}^{M^{(n)}}
  e_{-1}(\lambda_{i}^{(n)} - \lambda_{j}^{(n)})\ e_{-1}(\lambda_{i}^{(n)} +
  \lambda_{j}^{(n)}) e_{2}(\lambda_{i}^{(n)} - \lambda_{j}^{(n)})\
  e_{2}(\lambda_{i}^{(n)} + \lambda_{j}^{(n)}) \nonumber \\
  && \times \prod_{ j=1}^{M^{(n-1)}}e_{-1}(\lambda_{i}^{(n)} -
  \lambda_{j}^{(n-1)})\ e_{-1}(\lambda_{i}^{(n)} + 
  \lambda_{j}^{(n-1)}),
  \label{BAE1}
\end{eqnarray}
where we have introduced 
$$e_{x}(\lambda)=\frac{\lambda+\frac{ix}{2}}{\lambda-\frac{ix}{2}}\,.$$
It is interesting to note that equations (\ref{BAE1}) are exactly the
Bethe Ansatz equations of the $osp(1|\enne-1)$ case (see e.g. \cite{doikou1,
yabon}).

\subsubsection{$\bf sl(2n)$ algebra}

\begin{eqnarray}
  e_{1}(\lambda_{i}^{(1)})^{2L} &\!\!=\!\!& -\prod_{j=1}^{M^{(1)}}
  e_{2}(\lambda_{i}^{(1)} - \lambda_{j}^{(1)})\ e_{2}(\lambda_{i}^{(1)} +
  \lambda_{j}^{(1)})\ \prod_{ j=1}^{M^{(2)}}e_{-1}(\lambda_{i}^{(1)} -
  \lambda_{j}^{(2)})\ e_{-1}(\lambda_{i}^{(1)} + \lambda_{j}^{(2)})\,,
  \nonumber \\
  1 &\!\!=\!\!& -\prod_{j=1}^{M^{(l)}} e_{2}(\lambda_{i}^{(l)} -
  \lambda_{j}^{(l)})\ e_{2}(\lambda_{i}^{(l)} + \lambda_{j}^{(l)})\ \prod_{
  \tau = \pm 1}\prod_{ j=1}^{M^{(l+\tau)}}e_{-1}(\lambda_{i}^{(l)} -
  \lambda_{j}^{(l+\tau)})\ e_{-1}(\lambda_{i}^{(l)} +
  \lambda_{j}^{(l+\tau)}) \nonumber \\
  && l= 2,\ldots,n-1, \nonumber \\
  e_{-\theta_0}(\lambda_{i}^{(n)}) &\!\!=\!\!& -\prod_{j=1}^{M^{(n)}}
  e_{2}(\lambda_{i}^{(n)} - \lambda_{j}^{(n)})\ e_{2}(\lambda_{i}^{(n)} +
  \lambda_{j}^{(n)}) \nonumber \\
  && \times \prod_{ j=1}^{M^{(n-1)}}e_{-1}^{2}(\lambda_{i}^{(n)} -
  \lambda_{j}^{(n-1)})\ e_{-1}^{2}(\lambda_{i}^{(n)} + \lambda_{j}^{(n-1)}).
\label{BAE2}
\end{eqnarray}
The Bethe Ansatz equations are essentially the same as the ones obtained
from the \emph{folding} of the usual $sl(\enne)$ Bethe equations (see e.g.
\cite{dvgr, done}) for $M^{ (l)} =M^{(\enne-l)}$.
It can be realised from the study of the underlying
symmetry of the model that this \emph{folding} has
algebraic origins, as mentioned previously. 

\section{$sl({\emme}|{\enne})$ superalgebra\label{sect:slmn}}

In this section, we generalise the previous approach on (SP and SNP) 
boundary conditions to the $\ZZ_{2}$-graded case based on the
$sl({\emme}|{\enne})$ superalgebra. 
\subsection{Notations}
The $\ZZ_{2}$-gradation, depending on a sign $\theta_{0}=\pm$,  is
defined to be 
$(-1)^{[i]} = \theta_0$ for $i$ an $sl(\emme)$ index and 
$(-1)^{[i]} = -\theta_0$ an $sl(\enne)$ index.
\\
The $sl({\emme}|{\enne})$ invariant $R$ matrix reads
\begin{equation}
R_{12}(\lambda) = \lambda \II+iP_{12}\;,
\end{equation}
 where $P$ is from now on the super-permutation operator
(i.e. $X_{21}\equiv PX_{12}P$) such that
\begin{equation}
  \label{eq:Pdef}
  P = \sum_{i,j=1}^{\emme+\enne} (-1)^{[j]} E_{ij} \otimes E_{ji}
\end{equation}
The usual super-transposition  $^T$ 
is defined for any matrix 
$A = \sum_{ij} \; A^{ij} \;E_{ij}$, by
\begin{equation}
  \label{eq:st}
  A^{T} = \sum_{ij} (-1)^{[i][j]+[j]} \; A^{ji} \, E_{ij} = \sum_{ij}
  \left(A^T\right)^{ij} \, E_{ij} \;.
\end{equation}
As for the $sl(\enne)$ case, we will use a super-transposition $^t$
of the form 
\begin{equation}
  A^t=V^{-1}\,A^T\,V \;.
\end{equation}
The convention for $\theta_{0}$ and the expression of $V$ are chosen
accordingly to the selected Dynkin diagram.

Let us recall that for a basic Lie superalgebra, unlike the Lie 
algebraic
case, there exist in general many inequivalent simple root systems (i.e.
that are not related by a usual Weyl transformation), and hence many
inequivalent Dynkin diagrams. This situation occurs when a simple root
system contains at least one isotropic fermionic root. For each basic 
Lie
superalgebra, there is a particular Dynkin diagram which can be 
considered
as canonical: it contains exactly one fermionic root. Such a Dynkin 
diagram
is called distinguished. In the case of $sl(\emme|\enne)$, it has the
following form:
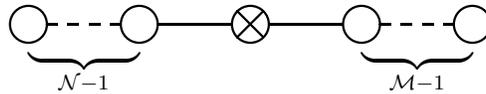
\setlength{\unitlength}{1pt}
\begin{center}
\begin{picture}(160,20) \thicklines
   \multiput(0,10)(42,0){5}{\circle{14}}
   \put(79,5){\line(1,1){10}}\put(79,15){\line(1,-1){10}}
   \put(7,10){\line(1,0){4}}\put(15,10){\line(1,0){4}}
   \put(23,10){\line(1,0){4}}\put(31,10){\line(1,0){4}}
   \put(49,10){\line(1,0){28}}
   \put(91,10){\line(1,0){28}}
   \put(133,10){\line(1,0){4}}\put(141,10){\line(1,0){4}}
   \put(149,10){\line(1,0){4}}\put(157,10){\line(1,0){4}}
   \put(0,0){$\underbrace{\hspace{42pt}}_{\enne-1}$}
   \put(126,0){$\underbrace{\hspace{42pt}}_{\emme-1}$}
\end{picture}
\end{center}
\vspace*{\baselineskip} In the case of $sl(\emme|2n)$ superalgebras, 
there
exists a symmetric Dynkin diagram with two isotropic fermionic simple 
roots
in positions $n$ and $\emme+n$:
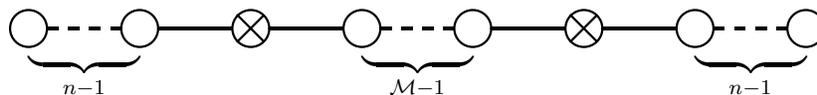
\begin{center}
\begin{picture}(280,20) \thicklines
   \multiput(0,10)(42,0){8}{\circle{14}}
   \put(79,5){\line(1,1){10}}\put(79,15){\line(1,-1){10}}
   \put(205,5){\line(1,1){10}}\put(205,15){\line(1,-1){10}}
   \put(7,10){\line(1,0){4}}\put(15,10){\line(1,0){4}}
   \put(23,10){\line(1,0){4}}\put(31,10){\line(1,0){4}}
   \put(49,10){\line(1,0){28}}
   \put(91,10){\line(1,0){28}}
   \put(133,10){\line(1,0){4}}\put(141,10){\line(1,0){4}}
   \put(149,10){\line(1,0){4}}\put(157,10){\line(1,0){4}}
   \put(175,10){\line(1,0){28}}
   \put(217,10){\line(1,0){28}}
   \put(259,10){\line(1,0){4}}\put(267,10){\line(1,0){4}}
   \put(275,10){\line(1,0){4}}\put(283,10){\line(1,0){4}}
   \put(0,0){$\underbrace{\hspace{42pt}}_{n-1}$}
   \put(126,0){$\underbrace{\hspace{42pt}}_{\emme-1}$}
   \put(252,0){$\underbrace{\hspace{42pt}}_{n-1}$}
\end{picture}
\end{center}
\vspace*{\baselineskip}
The generalization of the Weyl group for a basic Lie superalgebra gives 
a
method for constructing all the inequivalent simple root systems and 
hence
all the inequivalent Dynkin diagrams. For more
details, see \cite{Kac77,LSS86,DobrPetk,livre}.

\begin{description}
    \item[(i)] \textbf{Distinguished Dynkin diagram basis}\\
In this case, we consider that the $sl({\emme})$ part occupies the `upper' part
of the matrices and corresponds to bosonic degrees of freedom, whereas the
$sl({\enne})$ part occupies the `lower' part and corresponds to fermionic
degrees.  More precisely, the gradation takes the form:
\begin{eqnarray}
    (-1)^{[i]} = \begin{cases}\ 1 &\qmbox{for} 1\leq i\leq \emme \\
     -1 &\qmbox{for} \emme+1\leq i\leq \emme+\enne\,,\end{cases}
\end{eqnarray}
and the matrix $V$ reads
\begin{equation}
  V=\left(\begin{array}{c|c} V_\emme & 0 \\ 
      \hline 0 & V_\enne\end{array}\right)\,.
\end{equation}
In the above formula, $V_\emme$  (resp. $V_{\enne}$) is the $\emme\times\emme$ 
(resp. $\enne\times\enne$) matrix given in (\ref{eq:V}) for 
$\theta_{0}=+1$.
\item[(ii)] \textbf{Symmetric Dynkin diagram basis }\\
As in the $sl(\enne)$ case, one has to take $\emme$ or $\enne$ even. 
Note that for the
odd--odd case no symmetric Dynkin diagram exists and consequently no
twisted super-Yangian \cite{supYtw}. Here, we choose $\enne$ to be even:
$\enne=2n$. 
The $sl({\emme})$ part lies in the
`middle' part of the matrices and corresponds to fermionic degrees of
freedom, whereas the $sl({\enne})$ part occupies the `upper' and `lower'
part and is associated to bosonic degrees of freedom. Correlatively,
$\theta_0=-1$ in this case. The gradation 
is given by
\begin{eqnarray}
  (-1)^{[i]} = \begin{cases}\ 1 &\qmbox{for} 1\leq i\leq n\qmbox{and}
    \emme+n+1\leq  
    i \leq \emme+\enne\\
    -1 &\qmbox{for} n+1\leq i\leq \emme+n\,,\end{cases}
  \label{eq:indicessym}
\end{eqnarray}
while 
\begin{equation}
  V=\mbox{antidiag}\Big(\,
  \underbrace{1,\ldots,1}_{n+\emme}\,,\,\underbrace{-1,\ldots,-1}_{n}\,
  \Big)\,. 
\end{equation}
\end{description}
We will mostly use the distinguished Dynkin diagram basis in the
soliton preserving case, and the symmetric one in the soliton
non-preserving case.
In both cases, the $R$-matrix obeys the properties stated in section
\ref{sect:Rmatrix}, with 
$\bar R_{12}(\lambda)=R^{t_1}_{12}(-\lambda-i\rho)$ and $2\rho =
\theta_0(\emme -\enne)$.  The $K$-matrices will obey the defining 
relations stated in section \ref{sect:Kmatrix}, and the properties 
of the transfer matrix (section \ref{sect:transfer}) also hold; the
tensor product is now $\ZZ_{2}$-graded.

\subsection{Classification of reflection matrices for $\cY({\emme}|{\enne})$}

This section is devoted to the classification of reflection matrices for
the super-Yangian $\cY({\emme}|{\enne})$ based on $sl({\emme}|{\enne})$, both
for soliton preserving (prop. \ref{prop:SP1} and \ref{prop:SP2}) and for
soliton non-preserving boundary conditions (prop. \ref{prop:SNP1} and 
\ref{prop:SNP2}).

\subsubsection{Soliton preserving reflection}
\begin{proposition}
  \label{prop:SP1}
  Any bosonic invertible solution of the soliton preserving 
  reflection equation (RE)
  \begin{equation*}
    R_{12}(\lambda_{1}-\lambda_{2})K_{1}(\lambda_{1})
    R_{12}(\lambda_{1}+\lambda_{2}) K_{2}(\lambda_{2})=
    K_{2}(\lambda_{2})R_{12}(\lambda_{1}+\lambda_{2})
    K_{1}(\lambda_{1})R_{12}(\lambda_{1}-\lambda_{2})
  \end{equation*}
  where $R_{12}(\lambda)=\lambda\,\II+i\,P_{12}$ is the super-Yangian
  $R$-matrix, takes the form $K(\lambda) = U\,\left( i\xi\,\II
    +\lambda\,\EE\right)U^{-1}$ where either
  \begin{itemize}
  \item[(i)]
    $\EE$ is diagonal and $\EE^2=\II$ (diagonalisable solutions)
  \item[(ii)]
    $\EE$ is strictly triangular and $\EE^2=0$
    (non-diagonalisable solutions)
  \end{itemize}
  and $U$ is an element of the group $GL({\emme})\times GL({\enne})$. The
  classification is done up to multiplication by a function of the spectral
  parameter.
\end{proposition}
\textit{Proof}: Firstly it is obvious that for any solution $K(\lambda)$ to
the RE, and for any function $f(\lambda)$, the product
$f(\lambda)K(\lambda)$ is also a solution to the RE, so that the
classification will be done up to multiplication by a function of
$\lambda$.

\null

Expanding the reflection equation, one rewrites it as:
\begin{eqnarray}
  {[K_{2}(\lambda_{1}),\,K_{2}(\lambda_{2})]} &=&
  i(\lambda_{1}+\lambda_{2})\left( K_{2}(\lambda_{1})\,K_{1}(\lambda_{2}) -
  K_{2}(\lambda_{2})\,K_{1}(\lambda_{1})\right) \nonumber\\
  &&+i(\lambda_{1}-\lambda_{2})\left( K_{1}(\lambda_{1})\,K_{1}(\lambda_{2})
  - K_{2}(\lambda_{2})\,K_{2}(\lambda_{1})\right)
\end{eqnarray}
One then considers the RE with $\lambda_{1}$ and $\lambda_{2}$ exchanged,
and sums these two. After multiplication by $P_{12}$, one gets (for
$\lambda_{1}\neq\lambda_{2}$):
\begin{equation}
  {[K_{1}(\lambda_{1}),\,K_{1}(\lambda_{2})]} = -
  {[K_{2}(\lambda_{1}),\,K_{2}(\lambda_{2})]}
\end{equation}
the only solution of which is ${[K(\lambda_{1}),\,K(\lambda_{2})]} =0$. In other
words, the matrices $K(\lambda)$ at different values of $\lambda$'s are
diagonalisable (or triangularisable) in the same basis and must satisfy
\begin{equation}
  (\lambda_{1}+\lambda_{2})\left(K_{2}(\lambda_{1})\,K_{1}(\lambda_{2}) -
  K_{2}(\lambda_{2})\,K_{1}(\lambda_{1})\right)
  +(\lambda_{1}-\lambda_{2})\left(K_{1}(\lambda_{1})\,K_{1}(\lambda_{2}) -
  K_{2}(\lambda_{2})\,K_{2}(\lambda_{1})\right) = 0
\end{equation}
By setting $\lambda_{2}=-\lambda_{1}$, we get
$K(\lambda)K(-\lambda)=k(\lambda)\,\II$ for some function $k(\lambda)$. If
one now considers the case of invertible matrices, and since we are looking
for solutions up to a multiplicative function, we can take
$K(\lambda)K(-\lambda)=\II$, a condition which is generally assumed for
reflection matrices.

\null

We first consider the case where these matrices can be diagonalised:
$K(\lambda)=U\,D(\lambda)\,U^{-1}$, where $U$ is a group element of
$GL(m)\times GL(n)$. Projecting the RE on the basis element $E_{ii}\otimes
E_{jj}$, one gets
\begin{equation}
  (\lambda_{1}+\lambda_{2})\Big( d_{j}(\lambda_{1})\, d_{i}(\lambda_{2}) -
  d_{j}(\lambda_{2})\,d_{i}(\lambda_{1})\Big) =
  (\lambda_{1}-\lambda_{2})\Big( d_{j}(\lambda_{2})\,d_{j}(\lambda_{1}) -
  d_{i}(\lambda_{1}) \,d_{i}(\lambda_{2})\Big)
\end{equation}
where $D(\lambda)=\diag(d_{1}(\lambda),d_{2}(\lambda),\ldots,
d_{m+n}(\lambda))$. Since $K(\lambda)$ is supposed invertible, all the
$d_{j}$'s are not zero, and we consider
\begin{equation}
  q_{ij}(\lambda)=\frac{d_{i}(\lambda)}{d_{j}(\lambda)}
\end{equation}
which obeys
\begin{equation}
  (x+y)\Big(q(y)-q(x)\Big)+(x-y)\Big(q(x)q(y)-1\Big)=0\,. \label{eq-q}
\end{equation}
The solution to this equation is $q(x)=-\frac{x+i\xi}{x-i\xi}$ where $\xi$
is some complex parameter (including $\xi=\infty$), so that, considering
$q_{j1}(\lambda)$, we get
\begin{equation}
  d_{j}(\lambda)=-\frac{\lambda+i\xi_{j}}{\lambda-i\xi_{j}}\,d_{1}(\lambda)
  \ ,\qquad\forall\ j
\end{equation}
Requiring $q_{ij}(\lambda)$ to obey the equation (\ref{eq-q}) shows that one
must have
\begin{equation}
  d_{j}(\lambda)=\epsilon_{j}\lambda+i\xi \ \mbox{ with }\
  \epsilon_{j}=\pm1,\qquad \forall j
\end{equation}
where we have used the invariance under multiplication by a function. This
yields the form $(i)$, with $\EE=\diag(\epsilon_{1},\ldots,\epsilon_{m+n})$.

\null

We now turn to the case $K(\lambda)=U\, T(\lambda)\, U^{-1}$ where
$T(\lambda)$ is triangular. The projection of the RE on $E_{ii}\otimes
E_{jj}$ shows that the diagonal part of $T(\lambda)$ is still of the form
$(i)$. We distinguish two cases: $\EE$ has two different eigenvalues (which
are $\pm1$), or $\EE$ is proportional to $\II$ (and then the diagonal of
$T(\lambda)$ is also proportional to $\II$).

If $\EE$ has two eigenvalues, we project, in the first auxiliary space, on
two diagonal elements $E_{jj}$ and $E_{kk}$ associated to these eigenvalues:
\begin{equation}
  (\lambda_1+\lambda_2)\Big((i\xi\pm \lambda_2)T(\lambda_1) -(i\xi\pm
  \lambda_1)T(\lambda_2)\Big) = (\lambda_1-\lambda_2)\Big(
  T(\lambda_1)T(\lambda_2) -(i\xi\pm \lambda_1)(i\xi\pm \lambda_2)\Big)
\end{equation}
The difference and the sum of these equations read:
\begin{eqnarray}
  && \lambda_2\,T(\lambda_1)-\lambda_1\,T(\lambda_2) =
  \lambda_2-\lambda_1\,, \qquad\qquad \lambda_1+\lambda_2\neq0
  \label{diff}\\
  && i\xi(\lambda_1+\lambda_2)\Big(T(\lambda_1)-T(\lambda_2)\Big)=
  (\lambda_1-\lambda_2)\Big(T(\lambda_1)T(\lambda_2)+\xi^{2} -
  \lambda_1\lambda_2\Big) \qquad\label{som}
\end{eqnarray}
{}From equation (\ref{diff}), one gets
\begin{equation}
\frac{T(\lambda_1)-\II}{\lambda_1}=\frac{T(\lambda_2)-\II}{\lambda_2}
=T_{0}\,,\qquad i.e.\qquad T(\lambda)=\II+\lambda\,T_{0}
\end{equation}
where $T_{0}$ is a triangular matrix. Plugging this solution in equation
(\ref{som}), we obtain
\begin{equation}
  (i\xi-1)(\lambda_1+\lambda_2)\,T_{0}=\lambda_1\lambda_2(T_{0}^{2}-\II) +
  (\xi^{2}+ 1)\,\II, \qquad\forall \lambda_1,\lambda_2
\end{equation}
whose only (constant) solution is of the form $(i)$ with $i\xi=1$.

We are thus left with the case where the diagonal of $T(\lambda)$ is
proportional to the identity matrix: $T(\lambda)=\II+S(\lambda)$ with
$S(\lambda)$ strictly triangular. Projecting once more on a diagonal element
in the first auxiliary space, we obtain
\begin{eqnarray}
  && 2\Big(\lambda_2\,S(\lambda_1)-\lambda_1\,S(\lambda_2)\Big) =
  (\lambda_1-\lambda_2)S(\lambda_1)S(\lambda_2)\\
  &&\Leftrightarrow\ \frac{S(\lambda_1)}{\lambda_1(2\,\II+S(\lambda_1))} =
  \frac{S(\lambda_2)}{\lambda_2(2\,\II+S(\lambda_2))}=\sigma
\end{eqnarray}
where $\sigma$ is strictly triangular. We therefore have
$T(\lambda)=\II+2\lambda\,\sigma(\II-\lambda\sigma)^{-1}$. With this form
for $T(\lambda)$, the RE rewrites
$(\sigma_{1}-\sigma_{2})\sigma_{1}\sigma_{2}=0$, whose solution (for
strictly triangular matrices) is given by $\sigma^{2}=0$. Using this
property, we get the solution $(ii)$. \finproof

Note that the solutions given in this proposition are all of the form
$K(\lambda) = i\xi\,\II +\lambda\,\cE$ with $\cE^2=\II$ or $\cE^2=0$ (taking
$\cE = U \EE U^{-1}$).

\begin{proposition}
  \label{prop:SP2}
  Given a solution $K(\lambda) = i\xi\,\II +\lambda\,\cE$ to the
  soliton preserving RE  
\begin{equation}
  R_{12}(\lambda_{1}-\lambda_{2})\ K_{1}(\lambda_{1})\
  R_{21}(\lambda_{1}+\lambda_{2})\ K_{2}(\lambda_{2})=
  K_{2}(\lambda_{2})\
  R_{12}(\lambda_{1}+\lambda_{2})\ K_{1}(\lambda_{1})\
  R_{21}(\lambda_{1}-\lambda_{2})\;,
  \label{re-encore}
\end{equation}
and a
  solution $\bar K(\lambda) = i\xi'\,\II +\lambda\,\cE'$ to the
  anti-soliton preserving RE 
  identical to (\ref{re-encore}), the compatibility condition 
\begin{equation}
  \bar R_{12}(\lambda_{1}-\lambda_{2})\ K_{\bar 1}(\lambda_{1})\ \bar
  R_{21}(\lambda_{1}+\lambda_{2})\ K_{2}(\lambda_{2})= K_{2}(\lambda_{2})\
  \bar R_{12}(\lambda_{1}+\lambda_{2})\ K_{\bar 1}(\lambda_{1})\ \bar
  R_{21}(\lambda_{1}-\lambda_{2}) 
  \label{re3-encore}
\end{equation}
 is solved
  by $\cE' = \cE^t$ and $\xi+\xi'= \theta_0 \frac{\emme - \enne}2 \str \cE$.
\end{proposition}
\textit{Proof:} Straightforwardly, equation (\ref{re3-encore}) is equivalent to
\begin{eqnarray}
  \label{eq:EQE}
  {\cE'}^t_2 Q_{12} \cE_2 &=& \cE_2 Q_{12} {\cE'}^t_2 \\
  2 (\xi+\xi') [ \cE_2 , Q_{12} ] &=& [ \cE_2 , Q_{12} \cE_2 Q_{12} ]
\end{eqnarray}
The first equation yields $\cE' = \cE^t$. Using $Q_{12} \cE_2 Q_{12} =
\theta_0 \frac{\emme - \enne}2 Q_{12} \;\str \cE$ one gets the relation
between $\xi$ and~$\xi'$. \finproof

\subsubsection{Soliton non-preserving reflection}
\label{sect:class-SP}

\begin{proposition}
  \label{prop:SNP1}
  Any bosonic invertible solution of the soliton non-preserving RE 
  \begin{equation}
    R_{12}(\lambda_{1}-\lambda_{2})\ \tK_{1}(\lambda_{1})\ \bar
    R_{21}(\lambda_{1}+\lambda_{2})\ \tK_{2}(\lambda_{2}) =
    \tK_{2}(\lambda_{2})\ \bar R_{12}(\lambda_{1}+\lambda_{2})\
    \tK_{1}(\lambda_{1})\ R_{21}(\lambda_{1}-\lambda_{2}) 
\label{re2-encore}
\end{equation}
  where $R_{12}(\lambda)=\lambda\,\II+i\,P_{12}$ is the super-Yangian
  $R$-matrix, is a constant  
(up to a multiplication by a scalar function) matrix such that 
$\tK^t = \pm \tK$.
\end{proposition}
\textit{Proof}: Writing the $R$ and $\bar R$ matrices in terms of $\II$,
 $P_{12}$ and $Q_{12}$, and taking the part of (\ref{re2-encore}) 
which is symmetric in the exchange of
$\lambda_{1}$ and $\lambda_{2}$, yields the following equation
\begin{equation}
  \tK_{1}(\lambda_{1}) Q_{12} \tK_{1}(\lambda_{2}) +
  \tK_{1}(\lambda_{2}) Q_{12} 
  \tK_{1}(\lambda_{1}) = \tK_{2}(\lambda_{1}) Q_{12} K_{2}(\lambda_{2}) +
  \tK_{2}(\lambda_{2}) Q_{12} \tK_{2}(\lambda_{1})
  \label{eq:tmp1}
\end{equation}
In the same way, exchanging the role of spaces 1 and 2 from the original 
equation, one gets
\begin{equation}
  \tK_{1}(\lambda_{1}) Q_{12} \tK_{2}(\lambda_{2}) +
  \tK_{2}(\lambda_{1}) Q_{12} 
  \tK_{1}(\lambda_{2}) = \tK_{2}(\lambda_{2}) Q_{12} \tK_{1}(\lambda_{1}) +
  \tK_{1}(\lambda_{2}) Q_{12} \tK_{2}(\lambda_{1})
  \label{eq:tmp2}
\end{equation}
Transposing both equations (\ref{eq:tmp1}) and (\ref{eq:tmp2}) in space 1 and 
eliminating $P_{12}$, one gets after some algebra
\begin{equation}
  \tK^t(\lambda_{2}) =  f(\lambda_{1},\lambda_{2}) \tK(\lambda_{1}) \qquad 
  \forall \; \lambda_{1}, \lambda_{2}
\end{equation}
{}from which the final result follows.\finproof

\begin{proposition}
  \label{prop:SNP2}
  Given a solution $\tK_{1}$ to the soliton non-preserving RE 
\begin{equation}
  R_{12}(\lambda_{1}-\lambda_{2})\ \tK_{1}(\lambda_{1})\
  \bar R_{21}(\lambda_{1}+\lambda_{2})\ \tK_{2}(\lambda_{2})=
  \tK_{2}(\lambda_{2})\
  \bar R_{12}(\lambda_{1}+\lambda_{2})\ \tK_{1}(\lambda_{1})\
  R_{21}(\lambda_{1}-\lambda_{2})\;,
  \label{re2-encobis}
\end{equation}
and a
  solution $\tK_{\bar 1}$ to the CP-conjugate RE
  identical to (\ref{re2-encobis}), the compatibility condition 
\begin{equation}
  \bar R_{12}(\lambda_{1}-\lambda_{2})\ \tK_{\bar 1}(\lambda_{1})\ 
  R_{21}(\lambda_{1}+\lambda_{2})\ \tK_{2}(\lambda_{2})= \tK_{2}(\lambda_{2})\
  R_{12}(\lambda_{1}+\lambda_{2})\ \tK_{\bar 1}(\lambda_{1})\ \bar
  R_{21}(\lambda_{1}-\lambda_{2}) 
  \label{re3-encobis}
\end{equation}
 is solved by $\tK_{\bar 1}\propto\left(\tK_{1}\right)^{-1}$.
\end{proposition}
\textit{Proof:} Straightforward.\finproof

\subsection{Pseudo-vacuum and its dressing}

We can  determine explicitly
the eigenvalue $\Lambda_0(\lambda)$ of
the transfer matrix  
(defined as in section \ref{sect:transfer}) acting on the pseudo-vacuum 
$|\,\omega_{+}\rangle$ (which is always bosonic in our conventions).
We take here $K^{\pm}=\II$ (resp. $\tK^{\pm}=\II$), whilst cases with
non trivial $K^{\pm}$ (resp. $\tK^{\pm}$) are 
studied in section \ref{sect:SPK} (resp. \ref{sect:SNPK}).
$\Lambda_0(\lambda)$ is given by the following expression
\begin{equation}
  \label{eq:eigen0}
  \Lambda^{0}(\lambda) = \alpha(\lambda)^{L} g_{0}(\lambda) +
  \beta(\lambda)^{L} 
  \sum_{l=1}^{{\emme}+{\enne}-2} (-1)^{[l+1]}
  g_{l}(\lambda)
  + \gamma(\lambda)^{L} (-1)^{[{\emme}+{\enne}]}
  g_{{\emme}+{\enne}-1}(\lambda)
\end{equation}
where for (using the notation given in (\ref{abaBbB})):
\begin{description}
\item[(i) {Soliton preserving boundary conditions with $L$ sites}] (distinguished Dynkin diagram)
\begin{eqnarray}
  && \alpha(\lambda) = a^{2}(\lambda),
  ~~\beta(\lambda)=\gamma(\lambda)=b^{2}(\lambda)
\end{eqnarray}
and
\begin{eqnarray}
  \label{eq:glambda}
  && g_{l}(\lambda) = \frac{\lambda(\lambda + \frac{i({\emme}-{\enne})}{2})}
  {(\lambda + \frac{il}{2})(\lambda + \frac{i(l+1)}{2})} \;, \quad l =
  0,\ldots,{\emme}-1 \nonumber \\
  && g_{l}(\lambda) = \frac{\lambda(\lambda + \frac{i({\emme}-{\enne})}{2})}
  {(\lambda + \frac{i(2{\emme}-l-1)}{2})(\lambda + \frac{i(2{\emme}-l)}{2})}
  \;, \quad l = {\emme},\ldots,{\emme}+{\enne}-1
\end{eqnarray}
%$\rho = {{\emme} -{\enne} \over 2}$ in this convention. 
\\
\item[(ii) {Soliton non--preserving boundary conditions with $2L$ 
sites}] (symmetric Dynkin diagram)
\begin{eqnarray}
  && \alpha(\lambda) = \Big(a(\lambda) \bar b(\lambda)\Big)^2,
  ~~\beta(\lambda) = \Big(b(\lambda)\bar b(\lambda)\Big)^2,
  ~~ \gamma(\lambda) = \Big(\bar a(\lambda) b(\lambda)\Big)^2
\end{eqnarray}
and 
\begin{eqnarray}
  g_{l}(\lambda) &=& {\lambda + {i\over 2}(\rho +1) \over \lambda +
  {i\rho \over 2}}, \qquad 0\leq l<{{\emme} +{\enne} -1 \over 2}\nonumber \\
  g_{{\emme +{\enne}-1 \over 2}}(\lambda) &=&1,\qquad 
  \qmbox{if}\emme +{\enne} \qquad \mbox{odd}
  \nonumber \\ 
  g_{l}(\lambda) &=& g_{\enne+{\emme}-l-1}(-\lambda-i\rho). \qquad 
  \label{gb} 
\end{eqnarray}
We remind that $\theta_{0}=-1$ in that case.
\end{description}
%$\rho = {{\enne} -{\emme} \over 2}$ in this convention.
{From} the exact expression for the pseudo-vacuum eigenvalue, we introduce
the following assumption for the structure of the general eigenvalues:
\begin{eqnarray}
  \label{eq:eigen}
  \Lambda(\lambda) &=& \alpha(\lambda)^{L} g_{0}(\lambda) A_{0}(\lambda) +
  \beta(\lambda)^{L} \sum_{l=1}^{{\emme}+{\enne}-2} (-1)^{[l+1]}
  g_{l}(\lambda) A_{l}(\lambda) \nonumber\\
  &&+\;\gamma(\lambda)^{L} (-1)^{[{\emme}+{\enne}-1]}
  g_{{\emme}+{\enne}-1}(\lambda)
  A_{{\emme}+{\enne}-1}(\lambda) 
\end{eqnarray}
where the dressing functions $A_{i}(\lambda)$ need to be determined. 
The basic constraints that
they have to satisfy are the fusion and crossing equations as well as analyticity requirements.

\subsection{$sl({\emme}|{\enne})$ with soliton preserving boundary conditions}
\label{sect:slmnSP}

{From} the analyticity  of $\Lambda(\lambda)$,  one gets
\begin{eqnarray}
  \label{eq:anal1}
  A_{l}(-\frac{il}{2}) &=& A_{l-1}(-\frac{il}{2}), \qquad
  ~l=1,\ldots,{\emme}-1, \nonumber \\
  A_{2{\emme}-l}(-\frac{il}{2}) &=& A_{2{\emme}-l-1}(-\frac{il}{2}), \qquad
  l={\emme}-{\enne}+1,\ldots,{\emme}-1
\end{eqnarray}
Gathering the constraints (\ref{cross1}), (\ref{ft1}) and (\ref{eq:anal1}), one can
determine the dressing functions, i.e.
\begin{eqnarray}
  A_{0}(\lambda) &=& \prod_{j=1}^{M^{(1)}} {\lambda+
  \lambda_{j}^{(1)}-\frac{i}{2}\over \lambda+ \lambda_{j}^{(1)} +\frac{i}{2}
  }\ {\lambda-\lambda_{j}^{(1)}-{i\over 2} \over \lambda-\lambda_{j}^{(1)}
  +{i\over 2}} \nonumber \\
  A_{l}(\lambda) &=& \prod_{j=1}^{M^{(l)}}
  {\lambda+\lambda_{j}^{(l)}+{il\over 2}+i \over \lambda+ \lambda_{j}^{(l)}
  +{il\over2}} \; {\lambda-\lambda_{j}^{(l)}+{il\over 2}+i\over \lambda-
  \lambda_{j}^{(l)} +{il\over 2}} \nonumber \\
  && \times \prod_{j=1}^{M^{(l+1)}}{\lambda+ \lambda_{j}^{(l+1)}+{il\over
  2}-{i\over 2}\over \lambda+ \lambda_{j}^{(l+1)} +{il\over 2} +{i\over 2}}\
  {\lambda-\lambda_{j}^{(l+1)}+{il \over 2}-{i\over 2} \over
  \lambda-\lambda_{j}^{(l+1)} + {il\over 2}+{i\over 2}} \qquad l =
  1,\ldots,{\emme}-1 \nonumber \\
  A_{l}(\lambda) &=& \prod_{j=1}^{M^{(l)}}
  {\lambda+\lambda_{j}^{(l)}+i{\emme}-{il\over 2}-i \over \lambda+
  \lambda_{j}^{(l)} +i{\emme}-{il\over2}} \;
  {\lambda-\lambda_{j}^{(l)}+i{\emme}-{il\over 2}-i\over \lambda-
  \lambda_{j}^{(l)} +i{\emme}-{il\over 2}} \nonumber \\
  && \times \prod_{j=1}^{M^{(l+1)}}{\lambda+
  \lambda_{j}^{(l+1)}+i{\emme}-{il\over 2}+{i\over 2}\over \lambda+
  \lambda_{j}^{(l+1)} +i{\emme}-{il\over 2} -{i\over 2}}\
  {\lambda-\lambda_{j}^{(l+1)}+i{\emme}-{il \over 2}+{i\over 2} \over
  \lambda-\lambda_{j}^{(l+1)} +i{\emme}- {il\over 2}-{i\over 2}} \nonumber
  \\
  &&l = {\emme},\ldots,{\emme}+{\enne}-1
  \label{eq:dressingslmn}
\end{eqnarray}

\subsubsection{Bethe Ansatz equations for the distinguished Dynkin diagram}

{}From analyticity requirements one obtains the Bethe Ansatz equations,
\begin{eqnarray}
  e_{1}(\lambda_{i}^{(1)})^{2L} &\!\!=\!\!& -\prod_{j=1}^{M^{(1)}}
  e_{2}(\lambda_{i}^{(1)} - \lambda_{j}^{(1)})\ e_{2}(\lambda_{i}^{(1)} +
  \lambda_{j}^{(1)})\ \prod_{ j=1}^{M^{(2)}}e_{-1}(\lambda_{i}^{(1)} -
  \lambda_{j}^{(2)})\ e_{-1}(\lambda_{i}^{(1)} + \lambda_{j}^{(2)})\,,
  \nonumber \\
  1 &\!\!=\!\!& -\prod_{j=1}^{M^{(l)}} e_{2}(\lambda_{i}^{(l)} -
  \lambda_{j}^{(l)})\ e_{2}(\lambda_{i}^{(l)} + \lambda_{j}^{(l)})\ \prod_{
  \tau = \pm 1}\prod_{ j=1}^{M^{(l+\tau)}}e_{-1}(\lambda_{i}^{(l)} -
  \lambda_{j}^{(l+\tau)})\ e_{-1}(\lambda_{i}^{(l)} +
  \lambda_{j}^{(l+\tau)}) 
  \nonumber \\
  && l= 2,\ldots,{\emme}-1,{\emme}+1,\ldots,{\emme}+{\enne}-2 
  \nonumber \\
  1 &\!\!=\!\!& 
  \prod_{ j=1}^{M^{({\emme}-1)}}e_{-1}(\lambda_{i}^{({\emme})} -
  \lambda_{j}^{({\emme}-1)})\ e_{-1}(\lambda_{i}^{({\emme})} +
  \lambda_{j}^{({\emme}-1)}) 
  \nonumber \\
  &&\times 
  \prod_{j=1}^{M^{({\emme}+1)}} e_{1}(\lambda_{i}^{({\emme})}
  - \lambda_{j}^{({\emme}+1)})\ e_{1}(\lambda_{i}^{({\emme})} +
  \lambda_{j}^{({\emme}+1)})\ 
  \nonumber \\
  1 &\!\!=\!\!& -
  \prod_{j=1}^{M^{({\emme}+{\enne}-2)}}e_{-1}(\lambda_{i}^{({\emme}+{\enne}-1)}
  - \lambda_{j}^{({\emme}+{\enne}-2)})\
  e_{-1}(\lambda_{i}^{({\emme}+{\enne}-1)} +
  \lambda_{j}^{({\emme}+{\enne}-2)}) \nonumber \\
  &&\times \prod_{j=1}^{M^{({\emme}+{\enne}-1)}}
  e_{2}(\lambda_{i}^{({\emme}+{\enne}-1)} -
  \lambda_{j}^{({\emme}+{\enne}-1)})\
  e_{2}(\lambda_{i}^{({\emme}+{\enne}-1)} +
  \lambda_{j}^{({\emme}+{\enne}-1)})
  \label{BAE}
\end{eqnarray}
We recover here for $\emme=2$ and $\enne=1$ the Bethe Ansatz equation
of the supersymmetric $t-J$ model which corresponds to the $sl(2|1)$ 
case \cite{Foerster}.

\subsubsection{Bethe Ansatz equations for arbitrary Dynkin diagrams}
We wrote above only the dressing functions that correspond to the
distinguished Dynkin diagram.
It is however possible to construct the $g$ and dressing functions for
all the inequivalent Dynkin diagrams of $sl(\emme|\enne)$. 

The inequivalent Dynkin diagrams of the $sl(\emme|\enne)$ superalgebras
contain only bosonic root of same square length ("white dots"), usually
normalized to 2, and isotropic fermionic roots ("grey dots"). A given
diagram is completely characterized by the $p$-uple of integers
$0<n_{1}<\ldots<n_{p}<\emme+\enne$ labelling the positions of the grey
dots of the diagram.
Formally, we define $n_0=0$ and $n_{p+1}=\emme+\enne$ although there is
actually no root at these positions.
Such a diagram defined by the $p$-uple $(n_i)_{i=1\dots p}$
corresponds to the superalgebra $sl(\emme|\enne)$ with
\begin{equation}
  \emme = \sum_{\atopn{i \textrm{ odd}}{ i\leq p+1}} n_i
  - \sum_{i \textrm{ even} \atop i< p+1} n_i
  \qquad  \mbox{and}  \qquad
  \enne = \sum_{i \textrm{ even} \atop i\leq p+1} n_i
  - \sum_{i \textrm{ odd} \atop i< p+1} n_i \;.
\end{equation}
The $g$ functions have a form similar to
(\ref{eq:glambda}), with a change of increasing or decreasing
behaviour of the poles  each
time a grey (fermionic) root is met.
Indeed
\begin{eqnarray}
  \label{eq:glambdagen}
  && g_{l}(\lambda) = 
  \frac{\lambda\left(\lambda + \frac{i({\emme}-{\enne})}{2}\right)}
  {\left(\lambda + \frac{i}{2}\delta_l \right)
    \left(\lambda + \frac{i}{2}(\delta_l+1)\right)} \;, \quad l =
  0,\ldots,\emme+{\enne}-1 
\end{eqnarray}
where $\delta_0=0$ whilst the $\delta_l$ for $l=1,\dots,\emme+\enne-1$ are
obtained by iteration 
\begin{eqnarray}
    && \delta_l = 
  \begin{cases}
    \delta_{l-1} & \qmbox{if}  l=n_i \qmbox{for some } i \\
    \delta_{l-1}+1 & \qmbox{if}  n_{2i} < l < n_{2i+1} \qmbox{for some } i \\
    \delta_{l-1}-1 & \qmbox{if}  n_{2i-1} < l < n_{2i} \qmbox{for some } i 
  \end{cases}
\end{eqnarray}
The Bethe Ansatz equations read, for $\ell=1,\dots,\emme+\enne-1$ and
$i=1,\dots,M^{(\ell)}$ 
\begin{eqnarray}
  && 
  (1-\langle\alpha_\ell,\alpha_\ell\rangle) \,
  \prod_{k=1}^{\emme+\enne-1} \ 
  \prod_{j=1}^{M^{(k)}}
  e_{\langle\alpha_\ell,\alpha_k\rangle}(\lambda_{i}^{(\ell)} -
  \lambda_{j}^{(k)}) \ 
  e_{\langle\alpha_\ell,\alpha_k\rangle}(\lambda_{i}^{(\ell)} +
  \lambda_{j}^{(k)}) 
  = 
  \begin{cases}
    e_{1}(\lambda_{i}^{(1)})^{2L} & \ell=1\\[2mm]
    1 & \ell\neq 1
  \end{cases}
  \qquad\qquad
  \label{BAEgen}
\end{eqnarray}
where $\langle\alpha_\ell,\alpha_k\rangle$ is the scalar product of
the simple roots \emph{associated to the chosen Dynkin diagram}.

\subsubsection{Bethe Ansatz equations for the symmetric Dynkin diagram}

We give the useful example of the symmetric Dynkin
diagram for which  ${\enne}$ is even,
with the indices ordered as in (\ref{eq:indicessym}). The $g$
functions are in this case
\begin{eqnarray}
  \label{eq:glambdasym}
  && g_{l}(\lambda) = \frac{\lambda(\lambda + \frac{i({\emme}-{\enne})}{2})}
  {(\lambda + \frac{il}{2})(\lambda + \frac{i(l+1)}{2})} \;, \quad l =
  0,\ldots,{\enne}/2-1 \nonumber \\
  && g_{l}(\lambda) = \frac{\lambda(\lambda + \frac{i({\emme}-{\enne})}{2})}
  {(\lambda + \frac{i({\enne}-l-1)}{2})(\lambda + \frac{i({\enne}-l)}{2})}
  \;, \quad l = {\enne}/2,\ldots,{\emme}+{\enne}/2-1
  \nonumber \\
  && g_{l}(\lambda) = \frac{\lambda(\lambda + \frac{i({\emme}-{\enne})}{2})}
  {(\lambda + \frac{i(l-2{\emme})}{2})
    (\lambda + \frac{i(l-2{\emme}+1)}{2})}
  \;, \quad l = {\emme}+{\enne}/2,\ldots,{\emme}+{\enne}-1
\end{eqnarray}
and it is straightforward to get the $A_{i}$'s.
The Bethe Ansatz equations take the form:
\begin{eqnarray}
  e_{1}(\lambda_{i}^{(1)})^{2L} &\!\!=\!\!& -\prod_{j=1}^{M^{(1)}}
  e_{2}(\lambda_{i}^{(1)} - \lambda_{j}^{(1)})\ e_{2}(\lambda_{i}^{(1)} +
  \lambda_{j}^{(1)})\ \prod_{ j=1}^{M^{(2)}}e_{-1}(\lambda_{i}^{(1)} -
  \lambda_{j}^{(2)})\ e_{-1}(\lambda_{i}^{(1)} + \lambda_{j}^{(2)})\,,
  \nonumber \\
  1 &\!\!=\!\!& -\prod_{j=1}^{M^{(l)}} e_{2}(\lambda_{i}^{(l)} -
  \lambda_{j}^{(l)})\ e_{2}(\lambda_{i}^{(l)} + \lambda_{j}^{(l)})\ \prod_{
  \tau = \pm 1}\prod_{ j=1}^{M^{(l+\tau)}}e_{-1}(\lambda_{i}^{(l)} -
  \lambda_{j}^{(l+\tau)})\ e_{-1}(\lambda_{i}^{(l)} +
  \lambda_{j}^{(l+\tau)}) \nonumber \\
  && l= 2,\ldots,{\emme}+{\enne}-2, ~~l\neq {{\enne} \over 2},~{{\enne}
  \over 2}+{\emme} \nonumber \\
  1 &\!\!=\!\!& \prod_{j=1}^{M^{(l+1)}} e_{1}(\lambda_{i}^{(l)} -
  \lambda_{j}^{(l+1)})\ e_{1}(\lambda_{i}^{(l)} + \lambda_{j}^{(l+1)})\
  \prod_{ j=1}^{M^{(l-1)}}e_{-1}(\lambda_{i}^{(l)} - \lambda_{j}^{(l-1)})\
  e_{-1}(\lambda_{i}^{(l)} + \lambda_{j}^{(l-1)}) \nonumber \\
  && l= {{\enne} \over 2},~{{\enne} \over 2}+{\emme} \nonumber \\
  1 &\!\!=\!\!& -
  \prod_{j=1}^{M^{({\emme}+{\enne}-2)}}e_{-1}(\lambda_{i}^{({\emme}+{\enne}-
  1)} - \lambda_{j}^{({\emme}+{\enne}-2)})\
  e_{-1}(\lambda_{i}^{({\emme}+{\enne}-1)} +
  \lambda_{j}^{({\emme}+{\enne}-2)}) \nonumber \\
  && \times \prod_{j=1}^{M^{({\emme}+{\enne}-1)}}
  e_{2}(\lambda_{i}^{({\emme}+{\enne}-1)} -
  \lambda_{j}^{({\emme}+{\enne}-1)})\
  e_{2}(\lambda_{i}^{({\emme}+{\enne}-1)} +
  \lambda_{j}^{({\emme}+{\enne}-1)})
  \label{BAEsym}
\end{eqnarray}
The indices of the $e$'s in the products are the entries of the
Cartan matrix corresponding to the chosen Dynkin diagram, in
accordance with the results obtained in the closed chain case
\cite{saleur1, reshe, banania}.

\subsubsection{Non trivial soliton preserving boundary conditions}
\label{sect:SPK}

We come back to the distinguished Dynkin diagram basis and
 implement non trivial soliton preserving boundary conditions. From
the classification given in section \ref{sect:class-SP}, we know that
$K^-(\lambda)$
is always conjugated (by a constant matrix $U$) to a diagonal matrix of
the form
\begin{eqnarray}
  K(\lambda) = \diag( \underbrace{\alpha, \ldots ,\alpha}_{m_{1}},
  \underbrace{\beta, \dots, \beta}_{m_{2}}, \underbrace{\beta, \dots,
  \beta}_{n_{2}}, \underbrace{\alpha, \ldots ,\alpha}_{n_{1}} )
  \label{eq:solDiag}
\end{eqnarray}
As in the section \ref{sect:diagoK}, it is easy to see that the spectrum and the symmetry of
the model depend only on the diagonal (\ref{eq:solDiag}), and not on
$U$. Indeed, when considering two reflection matrices related by a
constant conjugation, the corresponding transfer matrices are also
conjugated. Thus, property \ref{prop:SP1} ensures that it is enough to
consider
diagonal $K(\lambda)$ matrices to get the general case. Such a
property, which relies on the form of the $R$-matrix, is a priori 
valid only in the \emph{rational} 
$sl(\enne)$ and $sl(\emme|{\enne})$ cases.
\\
For a diagonal solution (\ref{eq:solDiag})
with $m_{1}+m_{2}={\emme}$, $n_{1}+n_{2}={\enne}$, $\alpha(\lambda) =
-\lambda +i\xi$, $\beta(\lambda) = \lambda+i\xi$, and the free
boundary parameter $\xi$, one can compute the new form $\widetilde
g_{l}(\lambda)$  of the $g$-functions entering the expression of
$\widetilde\Lambda_{0}(\lambda)$, the new pseudo-vacuum eigenvalue.
They take the form:
\begin{eqnarray}
  \widetilde g_{l}(\lambda) &=& (-\lambda + i\xi)\, g_{l}(\lambda), \qquad
  l=0, \ldots, m_{1}-1 \nonumber \\
  \widetilde g_{l}(\lambda) &=& (\lambda + i\xi +im_{1})\, g_{l}(\lambda),
  \qquad l=m_{1}, \ldots, {\emme}+n_{2}-1 \nonumber \\
  \widetilde g_{l}(\lambda) &=& (-\lambda + i\xi - im_{2} + in_{2})\,
  g_{l}(\lambda), \qquad l={\emme}+n_{2}, \ldots, {\emme}+{\enne}-1
  \label{eq:tg1}
\end{eqnarray}
where $g_l(\lambda)$ are given by (\ref{eq:glambda}).
The dressing functions (\ref{eq:dressingslmn}) keep the same form, but
the Bethe Ansatz 
equations (\ref{BAE})  are modified (by $K^{-}(\lambda)$), so that the
value of the 
eigenvalues $\Lambda(\lambda)$ are different from the ones obtained
when $K(\lambda)=\II$.

\medskip

The modifications induced on Bethe Ansatz equations (\ref{BAE}) 
are the following: \\
-- The factor $-e_{2\xi+m_{1}}^{-1}(\lambda)$ appears in the LHS of the
${m_{1}}^{th}$ Bethe equation. \\
-- The factor $-e_{2\xi+m_{1}-m_{2}-n_{2}}^{-1}(\lambda)$ appears in the
LHS of the $({\emme}+n_{2})^{th}$ Bethe equation.

\subsection{$sl({\emme}|{\enne})$ with soliton non-preserving boundary conditions}

{}From equations of the type (\ref{fusiongen}) for the supersymmetric
 case relations
between the dressing functions are entailed for both soliton preserving and
soliton non-preserving boundary conditions. In particular, for the case that
corresponds to the symmetric Dynkin diagram one obtains (${\enne}=2n$, while
${\emme}$ can be even or odd),
\begin{eqnarray}
  \prod_{l=0}^{n-1} A_{l}(\lambda -il)\ \prod_{l=0}^{n-1}
  A_{{\emme}+n+l}(\lambda +i{\emme} -i(n-1) +il) =\prod_{l=0}^{{\emme}-1}
  A_{n+l}(\lambda -i(n-1) +il).
  \label{conge2}
\end{eqnarray}
In fact the latter equation is only necessary for the soliton non-preserving
boundary conditions.

\subsubsection{Dressing functions}  

As already mentioned we consider here the $R$ matrix, that corresponds to
the symmetric Dynkin diagram. Note that the $sl(2n|\emme)$ case is
isomorphic to $sl(\emme|2n)$ and entails the same dressing functions and BAE.

{}From the constraints (\ref{11}), (\ref{33}), (\ref{conge2}), we conclude
that the dressing functions take the form:
\begin{eqnarray}
  A_{0}(\lambda) &=& \prod_{j=1}^{M^{(1)}}{\lambda+
  \lambda_{j}^{(1)}-{i\over 2}\over \lambda+ \lambda_{j}^{(1)} +{i\over 2}}\
  {\lambda-\lambda_{j}^{(1)}-{i\over 2} \over \lambda-\lambda_{j}^{(1)}
  +{i\over 2}} \,, \nonumber\\
  A_{l}(\lambda) &=& \prod_{j=1}^{M^{(l)}}
  {\lambda+\lambda_{j}^{(l)}+{il\over 2}+i \over \lambda+ \lambda_{j}^{(l)}
  +{il\over2}} \; {\lambda-\lambda_{j}^{(l)}+{il\over 2}+i\over \lambda-
  \lambda_{j}^{(l)} +{il\over 2}} \nonumber \\
  && \times \prod_{j=1}^{M^{(l+1)}}{\lambda+ \lambda_{j}^{(l+1)}+{il\over
  2}-{i\over 2}\over \lambda+ \lambda_{j}^{(l+1)} +{il\over 2} +{i\over 2}}\
  {\lambda-\lambda_{j}^{(l+1)}+{il \over 2}-{i\over 2} \over
  \lambda-\lambda_{j}^{(l+1)} + {il\over 2}+{i\over 2}} \,, \qquad l =
  1,\ldots , n-1 \label{dressingSNPslmn}\\
  A_{l}(\lambda) &=& \prod_{j=1}^{M^{(l)}}
  {\lambda+\lambda_{j}^{(l)}+in-{il\over 2}-i \over \lambda+
  \lambda_{j}^{(l)} +in-{il\over2}} \;
  {\lambda-\lambda_{j}^{(l)}+in-{il\over 2}-i\over \lambda-
  \lambda_{j}^{(l)} +in-{il\over 2}} \nonumber \\
  && \times \prod_{j=1}^{M^{(l+1)}}{\lambda+
  \lambda_{j}^{(l+1)}+in-{il\over 2}+{i\over 2}\over \lambda+
  \lambda_{j}^{(l+1)} +in-{il\over 2} -{i\over 2}}\
  {\lambda-\lambda_{j}^{(l+1)}+in-{il \over 2}+{i\over 2} \over
  \lambda-\lambda_{j}^{(l+1)} +in- {il\over 2}-{i\over 2}} \,, \qquad
  n\le l < n+\frac{\emme-1}2 \nonumber
\end{eqnarray}
and $A_{l}(\lambda) = A_{\emme+2n-1-l}(-\lambda -i\rho)$,
and for $\emme=2m+1$
\begin{eqnarray}
  A_{k}(\lambda) &=& \prod_{j=1}^{M^{(k)}}
  {\lambda+\lambda_{j}^{(k)}+in-{ik\over 2}-i\over \lambda+
  \lambda_{j}^{(k)} +in-{ik\over2}} \;
  {\lambda-\lambda_{j}^{(k)}+in-{ik\over 2}-i\over \lambda-
  \lambda_{j}^{(k)} +in-{ik\over 2}} \nonumber \\ 
  && \times{\lambda+\lambda_{j}^{(k)}+in-{ik\over 2}+{i\over 2} \over
  \lambda+ \lambda_{j}^{(k)} +in-{ik\over 2}-{i\over 2}} \;
  {\lambda-\lambda_{j}^{(k)}+in-{ik\over 2}+{i\over 2}\over \lambda-
  \lambda_{j}^{(k)} +in-{ik\over 2}-{i\over 2}}\,, ~~k=m+n \label{a21}
\end{eqnarray}

Recall that the $sl(2m+1|2n+1)$ case is not examined because there is no
symmetric Dynkin diagram and consequently the `folding' of the algebra can
not be implemented. In any case it is known \cite{supYtw} 
that the twisted super-Yangian does not exist for $sl(2m+1|2n+1)$.

\subsubsection{Bethe Ansatz equations} 
 
{}From the analyticity requirements one obtains the Bethe Ansatz equations
which read as:
 
\subsubsection*{A. $\bf sl(2m+1|2n)$ superalgebra} 

\begin{eqnarray}
  e_{1}(\lambda_{i}^{(1)})^{2L} &\!\!=\!\!& -\prod_{j=1}^{M^{(1)}}
  e_{2}(\lambda_{i}^{(1)} - \lambda_{j}^{(1)})\ e_{2}(\lambda_{i}^{(1)} +
  \lambda_{j}^{(1)})\ \prod_{ j=1}^{M^{(2)}}e_{-1}(\lambda_{i}^{(1)} -
  \lambda_{j}^{(2)})\ e_{-1}(\lambda_{i}^{(1)} + \lambda_{j}^{(2)})\,,
  \nonumber \\
  1 &\!\!=\!\!&- \prod_{j=1}^{M^{(l)}} e_{2}(\lambda_{i}^{(l)} -
  \lambda_{j}^{(l)})\ e_{2}(\lambda_{i}^{(l)} + \lambda_{j}^{(l)})\ \prod_{
  \tau = \pm 1}\prod_{ j=1}^{M^{(l+\tau)}}e_{-1}(\lambda_{i}^{(l)} -
  \lambda_{j}^{(l+\tau)})\ e_{-1}(\lambda_{i}^{(l)} +
  \lambda_{j}^{(l+\tau)}) \nonumber \\
  && l= 2,\ldots,n+m-1, \;\; l \neq n \nonumber \\
  1 &\!\!=\!\!& \prod_{j=1}^{M^{(n+1)}} e_{1}(\lambda_{i}^{(n)} -
  \lambda_{j}^{(n+1)})\ e_{1}(\lambda_{i}^{(n)} + \lambda_{j}^{(n+1)})\
  \prod_{ j=1}^{M^{(n-1)}}e_{-1}(\lambda_{i}^{(n)} - \lambda_{j}^{(n-1)})\
  e_{-1}(\lambda_{i}^{(n)} + \lambda_{j}^{(n-1)}) \nonumber \\
  e_{-{1\over 2}}(\lambda_{i}^{(k)}) &\!\!=\!\!& -\prod_{j=1}^{M^{(k)}}
  e_{2}(\lambda_{i}^{(k)} - \lambda_{j}^{(k)})\
  e_{2}(\lambda_{i}^{(k)} + \lambda_{j}^{(k)})\ 
%%  \nonumber \\ && \times 
  e_{-1}(\lambda_{i}^{(k)} - \lambda_{j}^{(k)})\
  e_{-1}(\lambda_{i}^{(k)} + \lambda_{j}^{(k)})\ \nonumber \\
  && \times\prod_{j=1}^{M^{(k-1)}}e_{-1}(\lambda_{i}^{(k)} -
  \lambda_{j}^{(k-1)})\ e_{-1}(\lambda_{i}^{(k)} +
  \lambda_{j}^{(k-1)}) \,, ~~k=m+n 
  \label{BAE1n}
\end{eqnarray}

Note that the equations (\ref{BAE1n}) are the
Bethe Ansatz equations of the $osp(2m+1|2n)$ case (see e.g. \cite{yabon}) 
apart from the last equation.
 
\subsubsection*{B. $\bf sl(2m|2n)$ superalgebra} 
 
The first $n+m-1$ equations are the same as in the previous case see equation
(\ref{BAE1}), but the last equation is modified, with again  $k=m+n$,  to
\begin{eqnarray}
%%  e_{1}(\lambda_{i}^{(1)})^{2L} &\!\!=\!\!& -\prod_{j=1}^{M^{(1)}}
%%  e_{2}(\lambda_{i}^{(1)} - \lambda_{j}^{(1)})\ e_{2}(\lambda_{i}^{(1)} +
%%  \lambda_{j}^{(1)})\ \prod_{ j=1}^{M^{(2)}}e_{-1}(\lambda_{i}^{(1)} -
%%  \lambda_{j}^{(2)})\ e_{-1}(\lambda_{i}^{(1)} + \lambda_{j}^{(2)})\,,
%%  \nonumber \\
%%  1 &\!\!=\!\!& -\prod_{j=1}^{M^{(l)}} e_{2}(\lambda_{i}^{(l)} -
%%  \lambda_{j}^{(l)})\ e_{2}(\lambda_{i}^{(l)} + \lambda_{j}^{(l)})\ \prod_{
%%  \tau = \pm 1}\prod_{ j=1}^{M^{(l+\tau)}}e_{-1}(\lambda_{i}^{(l)} -
%%  \lambda_{j}^{(l+\tau)})\ e_{-1}(\lambda_{i}^{(l)} +
%%  \lambda_{j}^{(l+\tau)}) \nonumber \\
%%  && l= 2,\ldots,n+m-1, \;\; l \neq n \nonumber \\
%%  1 &\!\!=\!\!& \prod_{j=1}^{M^{(n+1)}} e_{1}(\lambda_{i}^{(n)} -
%%  \lambda_{j}^{(n+1)})\ e_{1}(\lambda_{i}^{(n)} + \lambda_{j}^{(n+1)})\
%%  \prod_{ j=1}^{M^{(n-1)}}e_{-1}(\lambda_{i}^{(n)} - \lambda_{j}^{(n-1)})\
%%  e_{-1}(\lambda_{i}^{(n)} + \lambda_{j}^{(n-1)}) \nonumber \\
  e_{1}(\lambda_{i}^{(k)}) &\!\!=\!\!& -\prod_{j=1}^{M^{(k)}}
  e_{2}(\lambda_{i}^{(k)} - \lambda_{j}^{(k)})\
  e_{2}(\lambda_{i}^{(k)} + \lambda_{j}^{(k)})\ 
%%  \nonumber \\ && \times
  \prod_{j=1}^{M^{(k-1)}}e_{-1}^2(\lambda_{i}^{(k)} -
  \lambda_{j}^{(k-1)})\ e_{-1}^2(\lambda_{i}^{(k)} +
  \lambda_{j}^{(k-1)}) \,. \qquad
  \label{BAE2n}
\end{eqnarray}
Notice that the Bethe Ansatz equations as in the non supersymmetric case,
are essentially the ones obtained from the \emph{folding} of the
\emph{symmetric} $sl({\emme}/{\enne})$ Bethe equations (\ref{BAEsym}) for
$M^{ (l)} =M^{({\emme}+{\enne}-l)}$. This \emph{folding} has algebraic
origins as can be realised from the study of the underlying symmetry of
the model (see (\ref{gen}), (\ref{gen2})). Indeed only half of the
$sl({\emme}|{\enne})$ generators survive after we impose the soliton
non-preserving boundary conditions, and these are exactly the generators of
the $osp({\emme}|{\enne})$ algebra.

\subsubsection{Non trivial soliton non-preserving boundary conditions}
\label{sect:SNPK}
We generalize the above approach to the diagonal case with
$\eps=1$ of the 
classification given in proposition \ref{prop:SNP1}: 
\begin{equation}
    \tK^{-}(\lambda)=\diag(k_{1},\ldots,k_{\emme+\enne})\qmbox{with}
    k_{\emme+\enne+1-j}=\,k_{j} \,.
\end{equation}
We will consider only invertible $\tK^{-}$ matrices, so that 
$k_i\neq 0 \ \forall i$.

The $g$-functions entering the new pseudo-vacuum eigenvalue are
modified in the following way: 
\begin{eqnarray}
  \widetilde g_{l}(\lambda) &=& k_{l+1}\, g_{l}(\lambda), \qquad
  0\leq l \leq \frac{\emme+\enne-1}2 
\end{eqnarray}
where $g_l(\lambda)$ are given by (\ref{gb}).
The remaining $\widetilde g$ are defined by requiring the
crossing relation
\begin{equation}
   \widetilde  g_{\emme+\enne-l}(-\lambda-i\rho)= \, 
   \widetilde  g_{l}(\lambda)\,.
\end{equation}
The dressing functions (\ref{dressingSNPslmn}) and (\ref{a21})
keep the same form, but the LHS of $\ell^{th}$ Bethe Ansatz
equation (given in (\ref{BAE1n}) and (\ref{BAE2n})) is multiplied by
$k_{\ell}/k_{\ell+1}$.

\vskip0.5cm

\textbf{Acknowledgements:} This work is supported by the TMR Network 
`EUCLID. Integrable models and applications: from strings to condensed
matter', contract number HPRN-CT-2002-00325.

\appendix

 \section{Survey of fusion}

We present here a brief review on fusion for systems with boundaries
\cite{mnf}. In particular we present the fusion procedure for open systems
without crossing symmetry developed in \cite{doikou1, doikou2}.

To cover both the SP and SNP cases, we define the action $*$ by: 
\begin{eqnarray}
  && R^*_{ab} = R_{ab},~~\bar R^*_{ab} = \bar R_{ab}, ~~K_{a}^* =
  K_{a}, ~~K_{\bar a}^* = K_{\bar a}, 
  ~~\mbox{for SP b.c.} \nonumber \\
  &&R^*_{ab} = \bar R_{ab},~~ \bar R^*_{ab} = R_{ab}, ~~K_{a}^* =
  \tK_{a},  ~~K_{\bar a}^* = \tK_{\bar a},
  ~~\mbox{for SNP b.c.} 
  \label{bc}
\end{eqnarray}
The $K_a^*$ and $K_{\bar a}^*$ 
matrices are solutions of the reflection (boundary Yang--Baxter)
equation
\cite{cherednik}
\begin{equation}
  R_{ab}(\lambda_{a}-\lambda_{b})\ K^*_{a}(\lambda_{a})\
  R^*_{ba}(\lambda_{a}+\lambda_{b})\ K^*_{b}(\lambda_{b})=
  K^*_{b}(\lambda_{b})\ R^*_{ab}(\lambda_{a}+\lambda_{b})\
  K^*_{a}(\lambda_{a})\ R_{ba}(\lambda_{a}-\lambda_{b}),
  \label{reetoile}
\end{equation} 
and they are related by the constraint
\begin{equation}
  \bar R_{ab}(\lambda_{a}-\lambda_{b})\ K^*_{\bar a}(\lambda_{a})\ \bar
  R^*_{ba}(\lambda_{a}+\lambda_{b})\ K^*_{b}(\lambda_{b})=
  K^*_{b}(\lambda_{b})\ 
  \bar R^*_{ab}(\lambda_{a}+\lambda_{b})\ K^*_{\bar a}(\lambda_{a})\ \bar
  R_{ba}(\lambda_{a}-\lambda_{b}) \;.
  \label{re3bis}
\end{equation}
These equations unify the equations (\ref{re})--(\ref{re3}) with
(\ref{re2})--(\ref{re4}). 

%% \begin{eqnarray}
%%   t^*(\lambda) =t(\lambda), ~~\mbox{for SP b.c.},~~t^*(\lambda) =\bar
%%   t(\lambda), ~~\mbox{for SNP b.c.}
%% \end{eqnarray}

The starting point for both cases (SP and SNP)
is the observation that the $\bar R$-matrix (\ref{cross}) at the
special value $\lambda = -i\rho$ yields a one-dimensional projector
onto the one-dimensional $sl(\enne)$-representation present in the
decomposition 
$ \enne\otimes \bar \enne=1 \oplus (\enne^{2}-1)$:
\begin{eqnarray}
  P_{\bar ab}^{-} = {1 \over \enne}\  \cP^{t_{a}}_{ab}
  \,=\, {1 \over \enne} \ Q_{ab} \; 
  = \; P_{a\bar b}^{-} \;.
%%   \qquad
%%   P_{\bar ba}^{-} = {1 \over \enne}\ V_{b}\ \cP^{t_{a}}_{ba}\ V_{b}^{-1}
%%   \,=\, {1 \over \enne} \ Q_{ba}\;.
\end{eqnarray}

This is related to the fact that the $\bar R$ matrix describes the
scattering between soliton and anti-soliton.  
Accordingly, the $({\enne}^{2}-1)$-dimensional projector is
\begin{eqnarray}
  P_{\bar ab}^{+} = 1 - P_{\bar ab}^{-} \label{pr2}.
\end{eqnarray}
Note that the soliton-soliton $R$ matrix at $\lambda
=-i$ provides a projector onto a $\enne$-dimensional space, reflected in
$\enne \otimes \enne = \enne\oplus (\enne^{2}-\enne)$. Thus, one
needs the $\bar R$ matrix, even in the SP case, hence the introduction of
$\bar t(\lambda)$ in both cases. 

We will formulate the fusion procedure for both types of boundary conditions
we mentioned. 

\medskip

We introduce the fused $R$-matrices
\begin{eqnarray}
  R_{<\bar ab>1}^*(\lambda) = P_{\bar ab}^{+}\ \bar R^*_{a 1}(\lambda)\
  R^*_{b1}(\lambda +i\rho)\ P_{\bar ab}^{+},\qquad R^*_{<b \bar a>1}(\lambda)
  = P_{b\bar a}^{+}\ R^*_{b 1}(\lambda)\ \bar R^*_{a 1}(\lambda +i\rho)\ P_{b
  \bar a}^{+},
\label{fr1}
\end{eqnarray}
and
\begin{eqnarray}
  R^*_{1<\bar ab>}(\lambda) = P_{\bar ab}^{+}\ R^*_{ 1b}(\lambda-i\rho)\
  \bar R^*_{1a}(\lambda )\ P_{\bar ab}^{+},\qquad R^*_{1<b \bar a>}(\lambda) =
  P_{b\bar a}^{+}\ \bar R^*_{ 1a}(\lambda-i\rho)\ R^*_{1b}(\lambda)\ P_{b \bar
  a}^{+}.
\label{fr2}
\end{eqnarray}
They satisfy generalised Yang-Baxter equations with fused indices.
Similarly, we use the reflection equation (\ref{reetoile}) and its
dual to obtain 
the fused $K$ matrices
\begin{eqnarray}
  K_{<\bar a b>}^{*-}(\lambda) &=& P_{\bar ab}^{+}\ K_{\bar
  a}^{*-}(\lambda)\ R^*_{b\bar a}(2\lambda+i\rho)\
  K_{b}^{*-}(\lambda+i\rho)\ P_{b\bar a}^{+}, \nonumber \\
  K_{<\bar a b>}^{*+}(\lambda) &=& P_{b\bar a}^{+}\ K_{\bar
  a}^{*+}(\lambda)\ R^*_{b\bar a}(-2\lambda-3i\rho)\
  K_{b}^{*+}(\lambda+i\rho)\ P_{\bar a b}^{+}.
\label{fk}
\end{eqnarray}
Both fused matrices obey generalised reflection equations of the type
(\ref{re3}) and its `dual' (for more details we refer the reader to
\cite{doikou2, mnf}). In an analogous way we obtain the $K^*_{<a \bar b>}$
matrices by fusing the spaces $a$ and $\bar b$. Now that the fused $R$ and
$K^*$ matrices are available, we operate the fusion of the transfer matrix
(\ref{t1}). The fused transfer matrix is defined by
\begin{eqnarray}
  t_F(\lambda)= \tr_{ab} \Big\{ K_{<\bar a b>}^{*+}(\lambda) \ T_{<\bar a
  b>}(\lambda) \ K_{<\bar a b>}^{*-}(\lambda) \ \hat T^*_{<b \bar
  a>}(\lambda+i\rho) \Big\} \;,
  \label{ft0}
\end{eqnarray}
where the fused $T$ matrices are obtained using (\ref{T1}) with fused $R$
matrices (\ref{fr1}) and (\ref{fr2}). After some algebra (see e.g.
\cite{mnf}) we end up with:
\begin{eqnarray}
  t_F(\lambda)= \zeta^*(2\lambda+2i\rho) \ \bar t(\lambda)\
  t(\lambda+i\rho) - \Delta[K^{*+}(\lambda)]\ \delta[T(\lambda)]\
  \Delta[K^{*-}(\lambda)]\ \delta[\hat T^*(\lambda)] \;,
  \label{ft1}
\end{eqnarray}
where $\zeta^*=\zeta$ and $\bar \zeta^*=\bar \zeta$ in the SP case, 
while $\zeta^*=\bar \zeta$ and $\bar \zeta^*=\zeta$ in the SNP case.
Furthermore the `quantum determinants' are (when we fuse the spaces $\bar a$
and $b$)
\begin{eqnarray}
  \delta[T(\lambda)] &=& \tr_{ab}\Big \{ P_{\bar a b}^{-}\ T_{\bar
  a}(\lambda)\ T_{b}(\lambda +i\rho)\Big \} \nonumber \\
  \delta[\hat T^*(\lambda)] &=& \tr_{ab}\Big \{P_{\bar a b}^{-} \ \hat
  T^*_{b}(\lambda)\ \hat T^*_{\bar a}(\lambda +i\rho)\Big \} \nonumber \\
  \Delta[K^{*-}(\lambda)] &=& \tr_{ab} \Big \{ P_{b \bar a }^{-}\ K_{\bar
  a}^{*-}(\lambda)\ R^*_{b\bar a}(2\lambda+i\rho)\
  K_{b}^{*-}(\lambda+i\rho)\Big \} \nonumber \\
  \Delta[K^{*+}(\lambda)] &=& \tr_{ab} \Big \{ P_{\bar a b}^{-}\
  K_{b}^{*+}(\lambda +i\rho)\ R^*_{\bar ab}(-2\lambda-3i\rho)\ K_{\bar
  a}^{*+}(\lambda) \Big \} \;.
  \label{qudet}
\end{eqnarray}
One obtains similar relations when the spaces $a$ and $\bar b$ are fused. To
compute the quantum determinants explicitly we use the following identities
which can be easily proved with the help of unitarity (\ref{uni1}) and the
crossing relation (\ref{cross})
\begin{align}
  & P_{\bar ab}^{-}\ R_{\bar a m}(\lambda)\ R_{bm}(\lambda +i\rho) =
  \zeta(\lambda+i\rho)\ P_{\bar ab}^{-} \;, && \\ 
  & P_{ a \bar b}^{-}\ R_{ a
    m}(\lambda)\ R_{\bar b m}(\lambda +i\rho) = 
  \bar\zeta(\lambda+i\rho)\ P_{ a\bar b}^{-} \;, 
  &&
  m=1, \ldots, L^{*} 
\end{align}
where $L^*=L$ in the SP case and $L^*=2L$ in the SNP case.
One then writes when fusing the spaces $\bar a$ and $b$
\begin{eqnarray}
  && \delta[T(\lambda)] = \zeta(\lambda+i\rho)^{L^{*}/2} 
  \zeta^*(\lambda+i\rho)^{L^{*}/2} \;,
  \qquad
 \delta[\hat T^*(\lambda)] = \zeta^*(\lambda+i\rho)^{L^{*}/2} 
 \zeta(\lambda+i\rho)^{L^{*}/2} \qquad
\end{eqnarray}
whilst, when we fuse the spaces $a$ and $\bar b$,
\begin{eqnarray}
  && \delta[T(\lambda)] = 
  \bar \zeta(\lambda+i\rho)^{L^{*}/2} 
  \bar \zeta^*(\lambda+i\rho)^{L^{*}/2} 
  \;,\qquad
  \delta[\hat T^*(\lambda)] =  \bar \zeta^*(\lambda+i\rho)^{L^{*}/2} 
  \bar \zeta(\lambda+i\rho)^{L^{*}/2} \;. \qquad
\end{eqnarray}
Furthermore, the r\^ole of $\zeta$ and $\bar\zeta$ is interchanged in
the latter 
equation depending whether the space $V_{m}$ belongs to the fundamental
representation or to its conjugate. This statement is important if one aims
at constructing the alternating spin chain. Finally for the special case
$K^{-} = 1$ and $K^{+} = 1$
\begin{eqnarray}
  \Delta[K^{*-}(\lambda)] =q^*(2\lambda +i\rho), \qquad
  \Delta[K^{*+}(\lambda)] =q^*(-2\lambda -3i\rho) \;,
\end{eqnarray}
where
\begin{eqnarray}
  q^*(\lambda) &=&q(\lambda) ~~\mbox{for SP}, ~~ q^*(\lambda) =\bar
  q(\lambda)~~\mbox{for SNP} \non\\ 
  q(\lambda)&=& \lambda -i\rho, ~~\bar q(\lambda)= \lambda +i.
\end{eqnarray} 

\section{Generalised fusion}

We describe a generalised fusion procedure for $sl({\enne})$ open spin
chains \cite{lepetit}. The procedure we use follows the lines of the
construction of 
the Sklyanin determinant for twisted Yangians \cite{MNO} and reflection
algebras \cite{momo,molev}.
The crucial observation here is that for the general case an one
dimensional projector can be also obtained by repeating the fusion procedure
${\enne}$ times, this is because ${\enne}^{ \otimes {\enne}} =
1\oplus...$ . The procedure described in the previous section 
is basically consequence of the fact that ${\enne} \otimes \bar {\enne} =
1\ \oplus \ ({\enne}^2-1)$.

Let us now introduce the following necessary objects for the
\emph{generalised} fusion procedure for open spin chains (see also
equations (2.13), (2.14) in \cite{avdo}),
\def\agras{{\boldsymbol{a}}}
\begin{eqnarray}
  T_{<\agras>} \equiv T_{<a_1...a_\enne>}= T_{a_{1}}(\lambda_{1}) \ldots
  T_{a_{{\enne}}}(\lambda_{{\enne}}),\qquad\qquad 
  \hat T^*_{<\agras>} = \hat
  T^*_{a_{1}}(\lambda_{1}) \ldots \hat T^*_{a_{{\enne}}}(\lambda_{{\enne}})
  \label{ftt} 
\end{eqnarray} 
where $\lambda_{l} =\lambda +i(l-1)$, $l=1 ,\ldots , \enne$ and $R^*$
defined in 
(\ref{bc}). For two sets $\{\mu_l\}_{l=1,\ldots ,\enne}$ and 
$\{\mu'_l\}_{l=1,\ldots ,\enne}$  we also define
\begin{eqnarray}
  {\cal R}^*_{<\agras>}(\{\mu_{l} \},\{\mu'_{l} \}) &=&
%%   R^*_{21}(\mu_{1} -\mu'_{2})R^*_{31}(\mu_{1} -\mu'_{3})
%%   \ldots R^*_{\enne 1}(\mu_{1} -\mu'_{\enne}) R^*_{32}(\mu_{2}
%%   -\mu'_{3}) \ldots R^*_{{\enne}2}(\mu_{2} -\mu'_{{\enne}}) \non\\
%%   && \ldots R^*_{k+1 k}(\mu_{k} -\mu'_{k+1}) \ldots
%%   R^*_{{\enne}k}(\mu_{k} -\mu'_{{\enne}}) \ldots R^*_{{\enne}
%%     {\enne}-1}(\mu_{{\enne}} -\mu'_{{\enne}-1}) \nonumber\\
%%  &=& 
  \prod_{k=1,...,\enne-1}^{\longrightarrow} 
  R^*_{a_{k+1} a_{k}}(\mu_{k} -\mu'_{k+1}) \ldots
  R^*_{a_{\enne}a_{k}}(\mu_{k} -\mu'_{{\enne}}) \;.
  \label{sol1} \label{frr} 
\end{eqnarray} 
In particular,
$\cR_{<\agras>}(\{\lambda_l\},\{\lambda_l\} )$ is proportional to the
antisymmetriser $\cA$, i.e. 
the projector onto a one-dimensional space 
(${\cal A}^2 = {\cal A}$). We also use
${\cal A}^+ =I-{\cal A}$ and
\begin{eqnarray}
  {\cal R}^{*+}_{<\agras>}(\{\mu_{l} \}) &\!=\!&
  \prod_{k=1,...,\enne-1}^{\longleftarrow} 
  R^*_{a_k a_{\enne}}(-\mu_{k} -\mu_{{\enne}}-2i\rho) \ldots
  R^*_{a_k a_{k+1}}(-\mu_{k} -\mu_{k+1}-2i\rho) 
  \label{frr2}
\end{eqnarray}

By multiplying the four equations 
\begin{alignat}{3}
  &
  {\cal R}^{*+}_{<\agras>} &\;  \equiv \;& 
  {\cal R}^{*+}_{<\agras>} (\{\lambda_l\}) \;=\; 
  {\cal A}\ {\cal R}^{*+}_{<\agras>}(\{\lambda_l\})
  + \cA^+\ {\cal R}^{*+}_{<\agras>}(\{\lambda_l\}) \;,
  \nonumber\\
  %% \qquad \qquad &&
  &
  T_{<\agras>} &\;=\;&
  {\cal A}\ T_{<\agras>} +\cA^+\ T_{<\agras>} \;, 
  \nonumber\\ 
  &
  {\cal R}^*_{<\agras>} &\;\equiv \;& 
  {\cal R}^*_{<\agras>} (\{\lambda_l\},\{-\lambda_l\}) \;=\; 
  {\cal A}\ {\cal R}^*_{<\agras>}(\{\lambda_l\},\{-\lambda_l\}) 
  + \cA^+\ {\cal R}^*_{<\agras>} (\{\lambda_l\},\{-\lambda_l\})
  \;,~~ 
  \nonumber\\
  &
  \hat T^*_{<\agras>} &\;=\;& {\cal A}\ \hat T^*_{<\agras>} +\cA^+\ \hat
  T^*_{<\agras>} 
  \label{four}
\end{alignat}
and keeping in mind that
\begin{eqnarray}
  {\cal A}\ {\cal R}^*_{<\agras>}\ {\cal A}^+ = {\cal A}\ {\cal
  R}^{*+}_{<\agras>}\ {\cal 
  A}^+ = {\cal A}\ T_{<\agras>} \ {\cal A}^+={\cal A}\ \hat
  T^*_{<\agras>} \ {\cal 
  A}^+ =0 
\end{eqnarray} 
we get
\begin{eqnarray}
  \tr_{<\agras>}( {\cal R}^{*+}_{<\agras>}\ T_{<\agras>}\ {\cal
  R}^*_{<\agras>}\ \hat T^*_{<\agras>}) &=& 
  \tr_{<\agras>}({\cal A}\ {\cal R}^{*+}_{<\agras>}\ {\cal A}\ \ T_{<\agras>}\
  {\cal A}\ {\cal 
  R}^*_{<\agras>}\ {\cal A}\ \hat T^*_{<\agras>}) \non\\
  &+& \tr_{<\agras>}({\cal A}^+\ {\cal R}^{*+}_{<\agras>}\ {\cal A}^+\
  \ T_{<\agras>}\ {\cal 
  A}^+\ {\cal R}^*_{<\agras>}\ {\cal A}^*\ \hat T^*_{<\agras>}). 
  \label{genf}
\end{eqnarray}  
Applying equation 
\begin{eqnarray}
  T_{a}(\lambda_{a})\ R^*_{ab}(\lambda_{a}+\lambda_{b})\ \hat
  T^*_{b}(\lambda_{b}) = \hat T^*_{b}(\lambda_{b})\
  R^*_{ab}(\lambda_{a}+\lambda_{b})\ T_{a}(\lambda_{a}) 
  \label{alg1}
\end{eqnarray} 
recursively we can show that 
\begin{equation}
  T_{<\agras>}\ {\cal R}^*_{<\agras>}\ \hat T^*_{<\agras>} = {\cal
  T}^*_{<\agras>}  
\end{equation}
where
\begin{eqnarray}
  &&{\cal T}^*_{<\agras>} = 
  \prod_{k=1,...,\enne}^{\longrightarrow} 
  \left(
    {\cal T}^*_{a_k} (\lambda_k)
    \prod_{l=k+1,...,\enne}^{\longrightarrow} 
    R^*_{a_l a_k}(\lambda_{l} +\lambda_{k}) 
  \right) \;.
  %%{\cal T}^*_{1} R^*_{21}(\lambda_{1}
  %%+\lambda_{2})R^*_{31}(\lambda_{1} +\lambda_{3}) \ldots
  %%R^*_{{\enne}1}(\lambda_{1} +\lambda_{{\enne}}) {\cal
  %%T}^*_{2}R^*_{32}(\lambda_{2} +\lambda_{3}) \ldots R_{{\enne}2}(\lambda_{2}
  %%+\lambda_{{\enne}}) \non\\ 
  %%&& {\cal T}^*_{3} \ldots {\cal T}^*_{k}R^*_{k+1 k}(\lambda_{k}
  %%+\lambda_{k+1}) \ldots R^*_{{\enne}k}(\lambda_{k}
  %%+\lambda_{{\enne}})\cT^*_{k+1} \ldots {\cal T}^*_{{\enne}-1} R^*_{{\enne}
  %%{\enne}-1}(\lambda_{{\enne}} +\lambda_{{\enne}-1}) {\cal T}^*_{{\enne}}.
  \label{sol1t} 
\end{eqnarray} 
However, as discussed in \cite{avdo} the trace of the above quantity
decouples to a product of ${\enne}$ transfer matrices, and therefore the LHS
of (\ref{genf}) simply becomes $\prod_{l=1}^{{\enne}}
t(\lambda_{l})$. Taking into account the property, 
\begin{eqnarray}
  {\cal A}\ {\cal O}_{<\agras>}\ {\cal A} = \tr_{<\agras>} ({\cal A}\
  {\cal O}_{<\agras>})\ 
  {\cal A}, 
\end{eqnarray} 
we can write the first term of the RHS of (\ref{genf}) as product of quantum
determinants, which are simply $c$ numbers, i.e.
\begin{eqnarray}
  \tr_{<\agras>}({\cal A}\ {\cal R}^{*+}_{<\agras>}\ {\cal A}\ \
  T_{<\agras>}\ {\cal A}\ {\cal 
  R}^*_{<\agras>}\ {\cal A}\ \hat T^*_{<\agras>})= \Delta\{K^{*+}(\lambda) \}\
  \delta\{T(\lambda)\}\ \Delta\{K^{*-}(\lambda) \}\ \delta\{\hat
  T^*(\lambda)\} 
\end{eqnarray} 
where
\begin{eqnarray}
  \Delta \{ K^{*+}(\lambda)\} &=& \tr_{<\agras>}\{{\cal A}\ {\cal
  R}^{*+}_{<\agras>}\},~~ 
  \delta\{T(\lambda)\}= \tr_{<\agras>}\{{\cal A}\ T_{<\agras>}\}, \non\\ 
  \Delta \{ K^{*-}(\lambda)\} &=& \tr_{<\agras>}\{{\cal A}\ {\cal
  R}^*_{<\agras>}\},~~\delta\{ \hat T^*(\lambda)\}=
  \tr_{<\agras>}\{{\cal A}\ \hat 
  T^*_{<\agras>}\}.
\end{eqnarray} 
Finally the second term of the RHS of (\ref{genf}) is simply the fused
transfer matrix $\tilde t(\lambda)$. Therefore, equation (\ref{genf}) can be
rewritten as
\begin{eqnarray}
  \tilde t(\lambda) = \prod_{l=1}^{{\enne}} t(\lambda_{l})
  -\Delta\{K^{*+}(\lambda) \}\ \delta\{T(\lambda)\}\
  \Delta\{K^{*-}(\lambda)\}\ 
  \delta\{\hat T^*(\lambda)\}. 
  \label{fusiongen}
\end{eqnarray}

\end{document}